\documentclass[prb,twocolumn,amsmath,amssymb,floats,showpacs,floatfix]{revtex4-2}
\usepackage{graphicx}
\usepackage{bm}
\usepackage{amsmath}
\usepackage{amsfonts}
\usepackage{amsbsy}
\usepackage{physics}
\usepackage{verbatim}
\usepackage{color}
\usepackage{soul}
\usepackage{float}
\usepackage{standalone}
\usepackage[splitrule]{footmisc}
\usepackage[hidelinks]{hyperref}
\usepackage[nameinlink]{cleveref}
\crefmultiformat{figure}{~#2#1#3}{,\!~#2#1#3}{, #2#1#3}{ and~#2#1#3}
\crefname{figure}{}{}
\usepackage{placeins}

\usepackage[caption=false]{subfig}
\usepackage{scalerel}
\usepackage[compat=1.0.0]{tikz-feynman}
\tikzfeynmanset{compat=1.0.0} 

\newcommand{\G}[4]{G^\text{#1}_{#2,\scaleto{#3}{4pt}}(#4)}

\DeclareMathOperator{\ii}{i}

\begin{document}

\title{Simulating electron-vibron energy transfer with quantum dots and resonators}
\author{C. Hermansen$^{1}$}
\thanks{These authors contributed equally to this work}
\author{M. Caltapanides$^{2}$}
\thanks{These authors contributed equally to this work}
\author{V. Meden$^{2}$}
\author{J. Paaske$^{1}$}

\affiliation{$^{1}$Center for Quantum Devices, Niels Bohr Institute, University of Copenhagen, 2100 Copenhagen, Denmark}
\affiliation{$^{2}$Institut für Theorie der Statistischen Physik, RWTH Aachen University and JARA—Fundamentals of Future Information Technology, 52056 Aachen, Germany}

\date{\today}
\begin{abstract}
Gateable semiconductor quantum dots (QDs) provide a versatile platform for analog quantum simulations of electronic many-body systems.
In particular, QD arrays offer a natural representation of the interacting $\pi$-electron system of small hydrocarbons. 
Here we investigate the prospects for extending QD simulators to encompass also the nuclear degrees of freedom. We represent the molecular vibrational modes by single-mode microwave resonators coupled capacitively to the QDs and study the gate-tunable energy transfer from a voltage-biased triple quantum dot (TQD) system to a single damped resonator mode.   
We determine the QD population inversions, the corresponding charge and energy currents as well as the resonator photon number, using Lindblad master equations and lowest-order perturbation theory within Keldysh Green function formalism. 
Along the way, we discuss the merits and shortcomings of the two methods.
A central result is the interrelation of a pronounced minimum in the charge current with a maximum in energy transfer, arising from a gate-tunable interference effect in the molecular orbitals of the TQD electron system. 
\end{abstract}

\maketitle

\section{Introduction}

The steady increase in complexity of gated quantum dot (QD) arrays has established this platform as a versatile tool for analog simulations of many-body physics. Due to the native Coulomb interaction, this platform comes with an obvious advantage for small-scale simulations of interacting electronic systems, which map directly onto extended Hubbard models. In GaAs/AlGaAs based QD arrays, this has been utilized to simulate small systems with tendencies towards a Mott-insulator instability~\cite{Hensgens2017Aug} and Nagaoka ferromagnetism~\cite{Dehollain2020Mar}, as well as tunable spin chains~\cite{vanDiepen2021Nov}. More recently, the formation of resonating valence bonds (RVB)~\cite{Wang2023Jun} and transport of excitons~\cite{Hsiao2024Mar} in Ge/SiGe based QD arrays were investigated. 

From an electronic perspective, small QD arrays with typically four to eight QDs constitute artificial molecules~\cite{QDMol2014}. More precisely, the $\pi$-electron systems in hydrocarbons are well described by a Pariser-Parr-Pople model for an array of $sp^{2}$-hybridized carbon $p_{z}$-orbitals~\cite{Pariser1953May, Pople1953Jan, Soos1984May}
\begin{align}
\label{eq:PPP}
    H_{\pi}=&\sum_{\langle ij\rangle,\sigma}t\hat{d}^{\dagger}_{i\sigma}\hat{d}^{}_{j\sigma}+\frac{1}{2}\sum_{ij,\sigma\sigma'}V_{ij}\hat{d}^{\dagger}_{i\sigma}\hat{d}^{\dagger}_{j\sigma'}\hat{d}^{}_{j\sigma'}\hat{d}^{}_{i\sigma},
\end{align}
where $\hat{d}^{(\dagger)}_{j\sigma}$ annihilates (creates) an electron in orbital $j$ with spin $\sigma$. 
The inter-carbon hopping integral and onsite Coulomb interaction are given by $t=-2.4\,$eV and $U=11.3\,$eV respectively. The latter decays with the distance $|\vec{r}_{ij}|$ between two $p_{z}$-orbitals, according to the Ohno representation~\cite{Ohno1964Jan} as $V_{ij}=U/\sqrt{1+|\vec{r}_{ij}|^{2}(U/14.4\,{\rm eV})^{2}}$, where $|\vec{r}_{ij}|$ is measured in {\AA}ngstr{\"o}m. Downscaling all energies by a factor of $10^{4}$, this is roughly matched by an appropriately tuned QD-array. In this sense, the array of four QDs of Ref.~\onlinecite{Wang2023Jun} , in which RVB states were observed, might already allow for simulation of the elusive cyclobutadiene molecule with four carbon atoms in a square~\cite{Senn1992Oct, Nakamura1989Sep}.

Here we propose an extension of the molecular QD-simulator to include also the nuclear degrees of freedom by representing the molecular vibrational modes by single-mode microwave resonators. Such QD-resonator hybrids are already available within the circuit quantum electrodynamics  (cQED) platform~\cite{Blais2021,Vigneau2023Jun}. Using high-impedance superconducting resonators strong couplings to a double quantum dot (DQD), i.e. larger than all dissipation rates, have been achieved~\cite{Frey2012Jan, Delbecq2011Dec, Toida2013Feb, Stockklauser2017Mar, FornDiaz2019Jun}. At present, the regimes of so-called ultrastrong, and deep strong couplings, on the order of a substantial fraction of, or larger than the resonator frequency, respectively, are beyond reach for quantum dots, but have been achieved for superconducting qubits~\cite{Yoshihara2017Jan, Gu2023May}. The available resonator frequencies ($f=\omega/2\pi$) typically range from a few tens of MHz to $10\,$GHz, which corresponds to a downscaling by a factor of $10^4$ of the infrared window ($10\,$GHz to $100\,$THz, i.e. $1-3000\,$cm$^{-1}$) for vibrational modes in real organic molecules.

In this sense, cooling down by the same factor of $10^{4}$ from room temperature, cryogenic QD-resonator arrays at $T=30\,$mK constitute simulators of small hydrocarbons, now including vibrational modes. Such devices would promote the previously studied QD-arrays~\cite{QDMol2014, Wang2023Jun, Hsiao2024Mar} to far more realistic artificial molecules, capable of emulating the rich interplay between vibrational and electronic degrees of freedom. 
This could facilitate simulations of phenomena like polaron formation, Jahn-Teller effects, conical intersections, isomerization, and possibly even aspects of reaction kinetics and photosynthesis~\cite{Volkhard2011, Ryndyk2015Dec}. 
With present-day QD-resonator devices used as molecular simulators, one could gain insight into the physics of parameter regimes that are otherwise difficult to study, e.g. due to large diabatic corrections beyond the Born-Oppenheimer approximation, poor separation of energy scales or strong nonlinearities. In terms of simulating model Hamiltonians, this extends the repertoire to include, for example, the paradigmatic and numerically challenging~\cite{Chen2016May, Wang2020Nov} Anderson-Holstein and Hubbard-Holstein models.

\begin{figure}[t]
    \centering
    \includegraphics[width=0.9\linewidth]{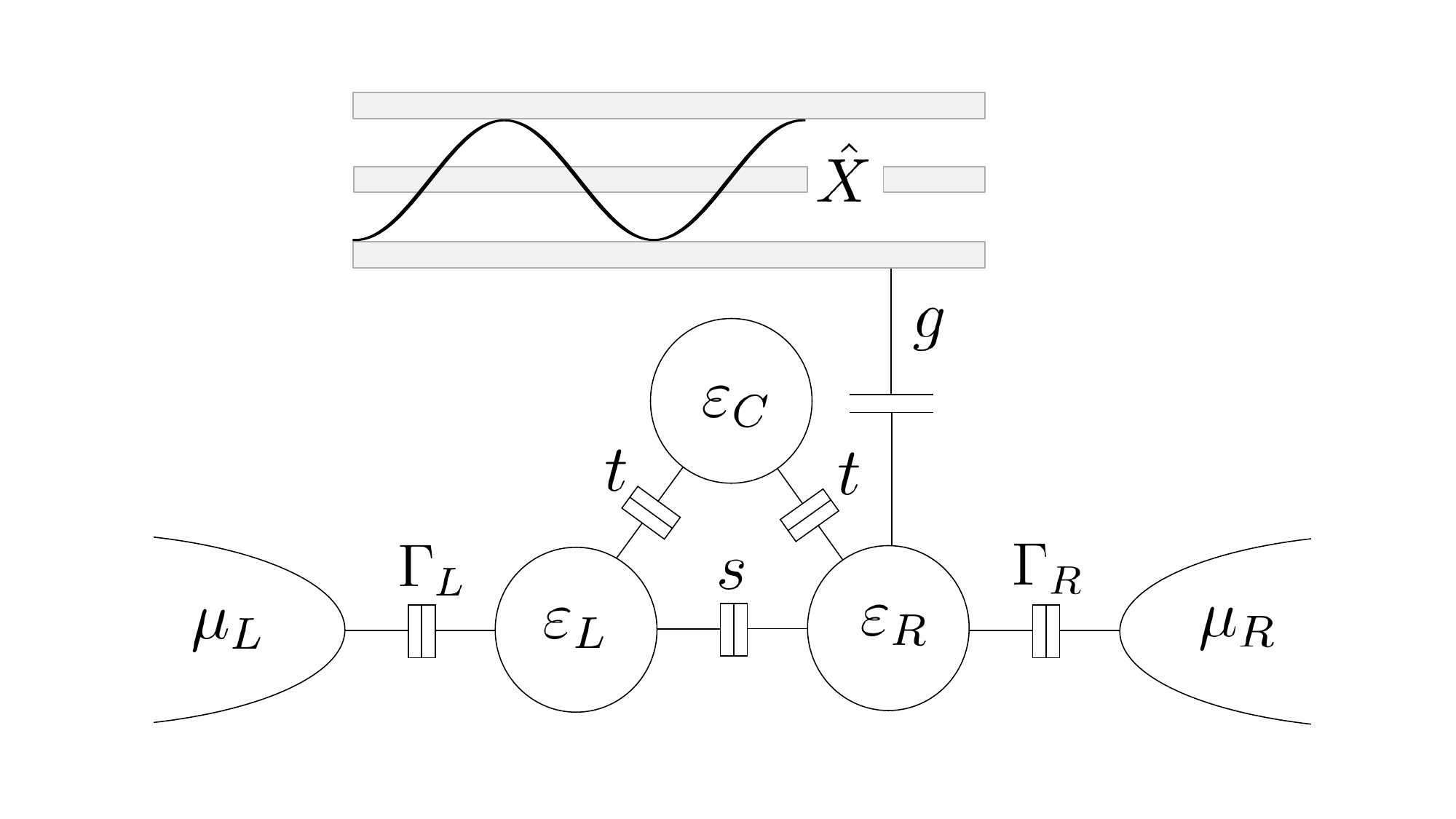}
    \caption{TQD-resonator hybrid system with inter-dot tunnel couplings $s$ and $ t$, which, by rates $\Gamma_{L/R}$, is coupled to two non-interacting metallic leads with different chemical potentials $\mu_{L/R}$. The dots onsite energies are given by $\varepsilon_{L/C/R}$. The right QD is capacitively coupled (amplitude $g$) to a microwave resonator which has a loss rate $\kappa$.
    We distinguish between a linear configuration (LTD), where $s=0$, and a triangular setup (TTD) with finite $s$.}
    \label{fig:model}
\end{figure}

Here, we analyze the steady-state energy transfer from a voltage-biased triple quantum dot (TQD) to a single microwave resonator (cf. Fig.\ref{fig:model}). As a simulator, this system constitutes an artificial single-molecule transistor~\cite{Park2000Sep, Evers2020Jul}, which, from an electronic point of view, is known to display I/V curves with Franck-Condon sidebands at voltages corresponding to integer multiples of the vibron frequency, and even Franck-Condon blockade of the sequential tunneling current for ultrastrong electron-vibron couplings~\cite{Braig2003Nov, Koch2005May, Burzuri2014Jun, Ryndyk2015Dec}. From the phonon perspective, which is challenging to study in real molecular transistors, the voltage-biased leads will provide a drive and a damping of the vibrational mode, which may lead to large current-induced energy transfer and possibly even a vibrational instability~\cite{Fedorets2002Apr, Hartle2011Mar, Lu2011Jul, Bode2011Jul, Erpenbeck2015May, Lu2019Feb}. 

The presence of such instabilities in molecules is analog to the onset of lasing in QD-resonator systems. Already a voltage-biased DQD coupled to a microwave cavity can function as a micromaser when gated such that electronic transitions become resonant with the cavity~\cite{Childress2004}. More precisely, the biased DQD serves as a gain medium, providing for resonant energy transfer into the cavity. Two DQDs in a microwave cavity have been demonstrated experimentally to give rise to a gain of the order of $10^3$ accompanied by a substantial line-narrowing of the cavity mode~\cite{Liu2015Jan, Liu2015Nov}. Although the criterion distinguishing nanolasers from light-emitting diodes is not entirely clear-cut~\cite{Saldutti2023May}, these basic traits of lasing clearly illustrate a resonant energy transfer from the electronic system to the microwave resonator. Conversely, the transfer of energy from photons to electrons in this DQD system has been shown to provide for efficient photon detection~\cite{Khan2021Aug, Haldar2024Feb}.
Theoretically, the voltage-biased DQD-resonator system has been thoroughly studied with an emphasis on lasing ~\cite{Childress2004, Jin2011Jul, Jin2012Nov, Agarwalla2016Sep}, photon statistics~\cite{Xu2013Nov, Zenelaj2022Nov}, photon detection~\cite{Wong2017Jan, Ghirri2020Jul} and generation of squeezed, and Schr\"{o}dinger cat states when including an additional ac drive to the cavity~\cite{Cottet2020Oct}. 

For the purpose of the present work, the TQD-resonator device approaches a more realistic molecular $\pi$-system, in the sense that the 'molecular orbitals' have more internal structure than for DQDs~\cite{Hsieh2012Oct}. Depending on the spatial layout of electrodes and resonators, orbital effects may lead to pronounced interferences~\cite{Pedersen2014Sep, Evers2020Jul, Chen2024Mar}.
Different TQD-resonator systems have already been fabricated and studied experimentally. Efficient gating between distinct charge states has been demonstrated using rf-reflectometry ~\cite{Braakman2013Mar, Braakman2013Jun, Delbecq2016Jan, Ansaloni2020Dec, Russ2020Jan}, and strong capacitive coupling between a resonant exchange qubit and a microwave resonator has been achieved~\cite{Medford2013Jul, Landig2018Aug, Landig2019Nov, Pan2020May}. 

Studying linear QD-chains of different lengths and with different resonator coupling configurations, it has been demonstrated that for certain setups, the TQD is a better photon emitter than both a DQD and a linear four-QD system~\cite{Agarwalla2016Sep}. 

With local gates on individual QDs, the linear TQD (LTD) (cf. Fig.~\ref{fig:model} for $s=0$) can be tuned to have two equidistant electronic transitions.
The triangular TQD (TTD) (cf. Fig.~\ref{fig:model} for $s\neq 0$), on the other hand, is the simplest system allowing for single-particle interference, which gives rise to a transmission node. In conjunction with the transmission node, the electron current exhibits a pronounced dip~\cite{Chen2015Feb, Rajput2023Feb}. Here, we couple the TQD to a resonator and demonstrate how the resonant energy transfer may be tuned by the local gates. We compare the LTD and the TTD configurations, and contrast the qualitatively different behaviors arising from their different 'molecular orbitals'. We focus mainly on the large-bias limit, where the bias voltage is larger than all other energy scales in the problem, and the energy transfer is most prominent.

The aim of this study is first and foremost to bring out a clear set of predictions for the resonator response to a large voltage-bias, much like it was measured in Ref.~\cite{Liu2015Jan}. Ideally, this is done alongside a measurement of the electron current, and with local gate-tunability of the individual dots and the next-nearest neighbor inter-dot tunneling barrier. The need for a molecular quantum simulator derives from the very fact, that the complexity of this nonequilibrium three-orbital Anderson Holstein model limits us severely in making reliable predictions. We therefore restrict our attention to certain manageable regions of parameter space, where calculations can be carried out. Our main simplification will be to restrict the full 64-dimensional electronic Hilbert space of the TQD to the 8-dimensional one of spin-polarized electrons. This approximation neglects all effects of on-site Coulomb interactions and circumvents Kondo correlations. We will discuss the relevance of this approximation towards the end of the paper.

We follow two complementary routes, using both second order perturbation theory for the self-energies in the QD-resonator coupling within the Keldysh Green function formalism as well as Lindblad master equations. A secondary purpose of this work is to compare these two methods and assess their range of validity, as well as individual advantages. 

This paper is organized as follows. In Sec.~\ref{sec:Model}, we introduce the model. Section~\ref{sec:QDspec} is devoted to a discussion of the basic features of the electronic many-body spectrum without the electron-photon interaction. Section~\ref{sec:method} outlines the methods used throughout, and Sec.~\ref{sec:ET} presents the results. In Sec.~\ref{sec:Coulomb}, we discuss the effects of Coulomb interactions and the implications of relaxing the spin-polarized approximation. We close with a discussion of our results and the perspectives for further molecular simulations using QD-arrays and microwave resonators. Supporting information and complementary results are relegated to appendices.  

\section{Model}
\label{sec:Model}

We consider a Holstein-like~\cite{Holstein1959Nov1, *Holstein1959Nov2} model given by
\begin{align}
    H&=H_\text{el}+H_\text{ph}+H_\text{el-ph}.
\label{eq:Hamiltonian}
\end{align}
The open electron system is described by a simple H\"{u}ckel model for the TQD system corresponding to the non-interacting part of Eq.~\eqref{eq:PPP}, with tunnel couplings to two metallic electrodes:
\begin{align}
    H_\text{el}&=\sum_{i=L,C,R}\varepsilon_i d_i^\dagger d_i 
    + \sum_{i=L,R;{\bf k}}\xi_{i{\bf k}}^{} c_{i{\bf k}}^\dagger c_{i{\bf k}}\nonumber\\
    &+ t(d_L^\dagger d_C + d_R^\dagger d_C + \text{h.c.}) + s(d_L^\dagger d_R+\text{h.c.})\nonumber\\
    &+\sum_{i=L,R;{\bf k}}\left(t_i c_{i{\bf k}}^\dagger d_i +\text{h.c.}\right).
\end{align}
Electrons on the three different QDs ($i=L,C,R$) with single-particle energies, $\varepsilon_i$, are created (annihilated) by $d_i^\dagger$ ($d_i$) and $t$ and $s$ are interdot tunnel couplings which we assume to be real and positive. Electrons with momentum ${\bf k}$ in the metallic leads, $i=L,R$, are created (annihilated) by $c_{i{\bf k}}^\dagger$ ($c_{i{\bf k}}$). Each lead is characterized by a featureless band, with dispersion $\xi_{i{\bf k}}=\varepsilon_{i{\bf k}}-\mu_{i}$, corresponding to a constant density of states, $\rho_{i}$. The two leads have chemical potentials, $\mu_{L/R}$, differing by an applied bias voltage, $V=\mu_{L}-\mu_{R}$, in units where $e=1$. They are assumed to be in thermal equilibrium at a temperature $T$, taken to be much smaller than all other energy scales in the problem.

The open resonator-bath system is described by the Hamiltonian
\begin{align}
    H_\text{ph}=&\,\omega_0 a^\dagger a  + \sum_s \omega_s a^\dagger_s a_s\nonumber\\
    &+\sum_s g_s (a^\dagger +a)(a_s^\dagger +a_s),
\label{eq:H_photon}
\end{align}
where microwave photons in the resonator with frequency $\omega_{0}$ are created (annihilated) by $a^\dagger$ ($a$).
The resonator is assumed to be linearly coupled by individual strengths, $g_{s}$, to a set of bosonic bath modes with frequencies $\omega_{s}$, created (annihilated) by $a_s^\dagger$ ($a_s$).

Finally, the electronic and photonic subsystems are assumed to interact locally by a resonator-induced shift of the potential on the right QD, 
\begin{align}
    H_\text{el-ph}&=gd_R^\dagger d_R(a^\dagger + a),
\label{eq:interaction}
\end{align}
with $g=\omega_{0}\alpha\sqrt{\pi Z_{0}/R_{K}}$ being the electron-photon (el-ph) coupling, set by the resonator impedance, $Z_{0}$, the resistance quantum, $R_{K}=h/e^{2}$, and the capacitive lever arm, $\alpha=C_{m}/(C_{m}+C_{g})$, given in terms of the mutual capacitance to the resonator $C_m$ and the total capacitance to ground of the right QD $C_g$~\cite{Childress2004}.

We focus on two different exemplary parameter sets, displaying two qualitatively different behaviors: The LTD and the TTD (cf. Fig.~\ref{fig:model}). For easy reference, these parameter sets are summarized in Table~\ref{tab:my_label}. 
In the majority of the paper we will be using these parameters, and any other choice of parameters will be stated explicitly.
\begin{table}[t]
\centering
\begin{tabular}{l c|c|c}\hline\hline
\textit{Parameters in units of t} & \hspace*{5mm}& \textbf{LTD}  & \textbf{TTD}  \\\hline\hline
       Next-nearest neighbor hopping $s$  & & 0 & 0.5  \\ \hline
       Left/right detuning $\varepsilon_L=-\varepsilon_R$ & &$\varepsilon/2$ & 0\\ \hline
       Central dot energy $\varepsilon_C$ & & $0$ & $\varepsilon_C$ \\ \hline
       Lead coupling rate $\Gamma_L=\Gamma_R= \Gamma$ & & \multicolumn{2}{c}{0.1}\\ \hline
       Temperature $T$ & & \multicolumn{2}{c}{$10^{-4}$}\\ \hline
       Voltage bias $\mu_L=-\mu_R=V/2$ & & \multicolumn{2}{c}{$10$} \\ \hline
       Resonator frequency $\omega_0$ & & \multicolumn{2}{c}{$3$} \\ \hline
       Resonator decay rate $\kappa$ & & \multicolumn{2}{c}{$0.005$}\\ \hline 
       Coupling strength $g$ & & \multicolumn{2}{c}{$0.1$} \\ \hline\hline
\end{tabular}
\caption{Model parameters in units of $t$. Unless otherwise stated these parameters are used in all plots below. System bath couling rates, $\Gamma$ and $\kappa$, are defined in Sec.~\ref{sec:method}}
\label{tab:my_label}
\end{table}

\section{The voltage-biased QD electron system}
\label{sec:QDspec}

\begin{figure}[t]
   \centering
    \includegraphics[width=0.49\linewidth]{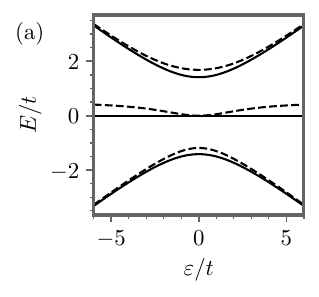}
    \includegraphics[width=0.49\linewidth]{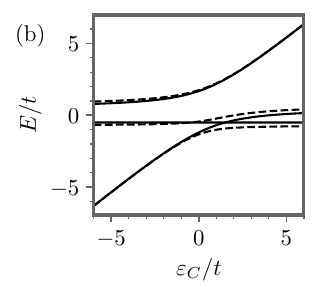}
    \caption{Single-particle spectrum. (a): LTD with $\varepsilon_C=0$  (full) or $\varepsilon_C=0.5t$ (dashed), (b): TTD with $\varepsilon=0$ (full) or $\varepsilon=t$ (dashed). }
   \label{fig:eigenenergies}
\end{figure}
In order to understand the energy transfer between electrons on the QDs and the photons in the resonator, it is illuminating to study first the isolated TQD system ($\Gamma=0$, $g=0$). The single-particle energies, obtained by straight-forward exact diagonalization of the TQD Hamiltonian, are plotted as a function of their respective level detunings in Fig.~\ref{fig:eigenenergies} for the LTD/TTD configurations. 

For the LTD at $\varepsilon_C=0$ (solid lines in Fig.~\ref{fig:eigenenergies} (a)), the eigenenergies are equidistant for all values of the left/right detuning, $\varepsilon$, so that two of the single-particle excitation energies are always degenerate and equal to $\Delta E=\sqrt{2t^2+\varepsilon^2/4}.$ A finite value of $\varepsilon_C$ splits this degeneracy (dashed lines in Fig.~\ref{fig:eigenenergies} (a)). For the TTD configuration (solid lines in Fig.~\ref{fig:eigenenergies} (b)) the two lowest-lying states exhibit a degeneracy for $\varepsilon_C=(t^2-s^2)/s=1.5t$, and therefore a degeneracy in the excitation energies $\Delta E = (2s^2+t^2)/s=3t$, which is lifted by a finite left/right detuning, $\varepsilon$ (dashed lines in Fig.~\ref{fig:eigenenergies} (b)).

\begin{figure}[t]
    \centering
    \includegraphics[width=\linewidth]{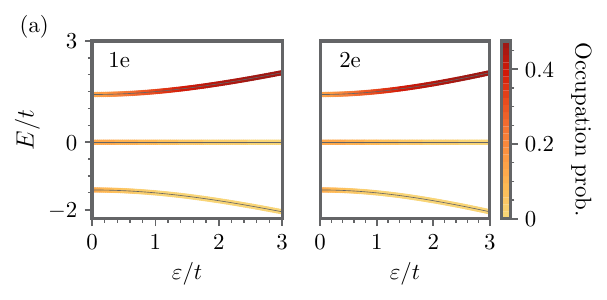}
    \includegraphics[width=\linewidth]{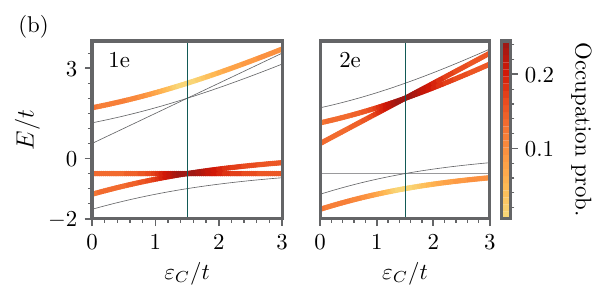}
    \caption{One- and two-electron spectra for LTD (a) and TTD (b) configurations, against left/right detuning and central dot energy, respectively. The coloring of the lines indicates the occupation probability. For the 1e/2e charge sector, thin full lines show the spectrum of the 2e/1e charge state. 
    The gridlines at $\varepsilon_{C}=1.5 t$ in (b) are shown for easy reference to Fig.~\ref{fig:spectral_filling}. Note that the occupations do not sum to one as also the zero- and three-electron states (not shown) acquire a finite occupation probability}
    \label{fig:eigenstate_occ_tri}
\end{figure}

Electronic population inversion induced by a large bias voltage will play a central role for the energy transfer. The population inversion of the many-body eigenstates at infinite bias voltage 
is therefore displayed in Fig.~\ref{fig:eigenstate_occ_tri}, computed using Lindblad master equations as described below in Sec.~\ref{sec:Lindblad}. It is important to note that the eigenenergies plotted are for the isolated TQD without coupling to leads or resonator ($\Gamma=0, g=0$), whereas the occupation probabilities are those of the closed-system eigenstates calculated using the steady-state density matrix for the open system at inifite bias voltage ($\Gamma\neq 0, g=0$). The true eigenstates of the coupled system no longer have zero linewidth as in Fig.~\ref{fig:eigenstate_occ_tri} but acquire a finite lifetime broadening due to hybridization with the leads. The one- and two-electron (1e/2e) spectra are identical for the LTD configuration, whereas in the 2e spectrum for the TTD, it is the two excited states which are degenerate at $\varepsilon_{C}=1.5 t$ (Fig.~\ref{fig:eigenstate_occ_tri} (b)).
We note that the LTD exhibits population inversion with large occupation of the highest excited state for a wide range of detunings, whereas the TTD only shows population inversion among the 2e-states close to the degeneracy point. As $H_\text{el-ph}$ is charge conserving, we only need to consider population inversion within the 1e/2e charge sector.

\begin{figure}[t]
    \centering
    \includegraphics[width=0.49\linewidth]{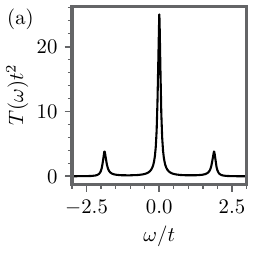}
    \includegraphics[width=0.49\linewidth]{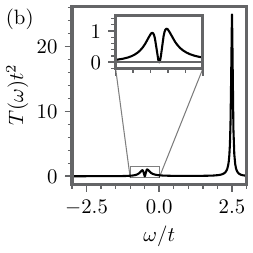}
    \includegraphics[width=0.49\linewidth]{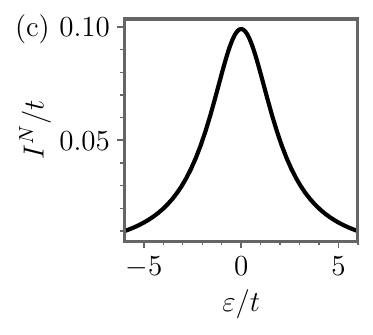}
    \includegraphics[width=0.49\linewidth]{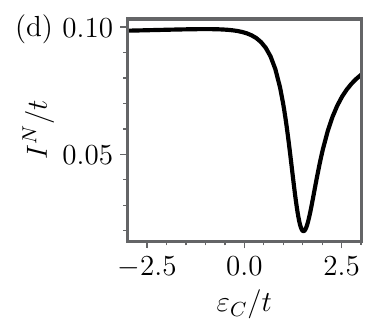}
     \caption{Transmission function $T(\omega)=|G^\text{R}_{RL}(\omega)|^2$ ((a)-(b)) and electron current ((c)-(d)) for  the uncoupled ($g=0$) TQD system.
     (a)/(c): LTD with $\varepsilon=2.5t$, (b)/(d): TTD with $\varepsilon_C=1.5t$.}
    \label{fig:Transmission_function}
\end{figure}

Further insight into the dynamics of the electron system can be obtained from the frequency-dependent left-right transmission functions $T(\omega)=|G^\text{R}_{RL}(\omega)|^2$, with $G^\text{R}_{RL}(\omega)$ being the right-left matrix element of the retarded Green function, shown in Fig.~\ref{fig:Transmission_function} (a)-(b). They were computed using the Green function formalism as described in Sec.~\ref{sec:Keldysh}.
The LTD displays a large zero-frequency transmission peak together with two smaller satellites appearing at the single-particle excitation energies in Fig.~\ref{fig:eigenenergies} (a).
The TTD, on the other hand, shows a transmission peak at the highest single-particle excitation energy, and an interference node at the lowest one (at the degeneracy point). This transmission node is caused by an interference in the mirror-symmetric TTD configuration. It blocks transport from the left to the right dot, while electrons on the right dot can easily escape into the lead, resulting in the pronounced population inversion among the 2e states at $\varepsilon_{C}=1.5 t$ (compare Fig.~\ref{fig:eigenstate_occ_tri} (b)). This effect is explained in greater detail in Sec.~\ref{sec:Population_Inversion}.
As shown earlier in Ref.~\cite{Chen2015Feb}, these transmission properties are reflected in the electron current, plotted in Fig.~\ref{fig:Transmission_function} (c)-(d). Whereas the LTD configuration displays a peak at the charge-degeneracy point, $\epsilon=0$, the transmission node leads to a pronounced current minimum at the degeneracy point, $\varepsilon_{C}=1.5 t$, in the TTD configuration.

This summarizes the salient features of the voltage-biased, non-interacting ($g=0$) electron system, calculated as detailed below. When considered as a potential gain medium, it is evident that both systems display substantial population inversion. The LTD has two distinct electronic transitions within both the 1e and 2e sectors, but only one of them is population inverted (see Fig. \ref{fig:eigenstate_occ_tri} (a)). The TTD can be tuned to a transmission node, where two population inverted transitions are available (see Fig. \ref{fig:eigenstate_occ_tri} (b)). 

\section{Methods}
\label{sec:method}

To investigate the steady state of the bias voltage-driven TQD coupled to a lossy resonator we employ two complementary methods: perturbation theory within the Keldysh Green function formalism and the Lindblad master equation approach.

\subsection{Keldysh Green functions}
\label{sec:Keldysh}

Starting from the Hamiltonian, Eqs.~(\ref{eq:Hamiltonian})-(\ref{eq:interaction}), the non-equilibrium many-body problem can conveniently be formulated by means of a  Keldysh path integral ~\cite{Kamenev2023Jan}. The partition function
\begin{equation}
    Z=\int\mathcal{D}[\bar{\phi},\phi, \bar{\psi},\psi]e^{i\left(S_{\text{el}}[\bar{\psi},\psi]+S_\text{ph}[\bar{\phi},\phi]+S_\text{el-ph}[\bar{\phi},\phi, \bar{\psi},\psi]\right)},
    \label{eq:partition_function}
\end{equation}
can be expressed in terms of a Keldysh action $S = S_{\text{el}} + S_{\text{ph}} + S_{\text{el-ph}}$.
Here the Grassmann fields $\psi$ refer to QD electrons and the complex fields $\phi$ to the resonator photons. 

The electronic part of the action is given by
\begin{align}
    S_\text{el}=\int^\infty_{-\infty} \dd t \dd t'\, \bar{\Psi}_D(t)\check{G}_0^{-1}(t,t')\Psi_D(t')   
\end{align}
with $\Psi_D=(\psi_{1L},\, \psi_{1C},\,\psi_{1R},\,\psi_{2L},\,\psi_{2C},\,\psi_{2R})^T$, 
where $1$ and $2$ denote the Keldysh rotated contour indices. The bare Green function is a matrix in Keldysh and dot space
\begin{align}
    \check{G}_{0}
    =\begin{pmatrix} 
        G_{0}^\text{R}(t,t') && G_{0}^\text{K}(t,t') \\
        0 && G_{0}^\text{A}(t,t')
    \end{pmatrix}.
\end{align}
In the steady-state, it is obtained from the Dyson equation with exact tunneling self-energies given by
\begin{align}
    &\Sigma^\text{R/A}_\text{bath,f}(\omega)=\mp i\Gamma\begin{pmatrix}
    1&&0&&0\\
    0&&0&&0\\
    0&&0&&1\end{pmatrix}\nonumber\\
    &\Sigma^\text{K}_\text{bath,f}(\omega)=-2i\Gamma\begin{pmatrix}
    F^{(0)}_{L}(\omega) && 0&&0\\0&&0&&0\\0&&0&& F^{(0)}_{R}(\omega)
    \end{pmatrix}
\label{eq:GKinv}
\end{align}
in frequency space. We have employed the wide-band limit with frequency-independent tunneling rates, $\Gamma_i=\pi\rho_{i}|t_i|^2$, resulting from constant density of states in the leads. For simplicity, we assume symmetric couplings $\Gamma_L=\Gamma_R=\Gamma$. Each lead is assumed to be in thermal equilibrium at a temperature $T$, taken to be much smaller than any other energy scale in the problem. The Keldysh component of the Green function is expressed in terms of the equilibrium lead distribution function, $F^{(0)}_{i}(\omega)=\tanh[(\omega-\mu_{i})/2T]$, in units where $k_{B}=1$.
 
The photon part of the action can be written in terms of the real (dimensionless) fields $X$ and $P$. A Gaussian integration over the momentum $P$ can be performed to obtain an effective action for the oscillator displacement $X$. The resonator mode is coupled to a bosonic bath with ohmic density of states $J(\omega)=4\kappa\omega$, 
assumed to be in 
thermal equilibrium with the same temperature, $T$, as the electrons. The bilinear form of the resonator-bath interaction in Eq.~\eqref{eq:H_photon} allows for straightforward Gaussian integration over the bath fields.
The resulting photon self-energies (bosonic bath) take the form
\begin{align}
    \Sigma^\text{R/A}_\text{bath,b}(\omega)&=\mp 2i \kappa \omega/\omega_0,\\
    \Sigma^\text{K}_\text{bath,b}(\omega)&=\left(\Sigma^\text{R}_\text{bath,b}-\Sigma^\text{A}_\text{bath,b}\right)\coth\left(\frac{\omega}{2T}\right),
\end{align}
and the corresponding effective photon action now becomes
\begin{align}
    &S_\text{ph}
    =\frac{1}{2}\int \frac{\dd\omega}{2\pi}\, 
    \left(X^{{\rm cl}}\,\,X^{\rm q}\right)_{-\omega}
    \check{D}_0^{-1}(\omega)
    \begin{pmatrix}X^{{\rm cl}}\\X^{\rm q}
    \end{pmatrix}_{\!\!\omega}\label{eq:bare_photon_action},
\end{align}
with retarded/advanced Green functions given by
\begin{equation}
    D^\text{R/A}_{0}(\omega)=\frac{\omega_0}{\omega^2-\omega_0^2\pm 2i\kappa\omega}.
    \label{eq:retarded_photon}
\end{equation}
and the quantum (q) and the classical (cl) parts of the displacement $X^{\rm q/cl}$~\cite{Kamenev2023Jan}. 
The Keldysh Green function satisfies the fluctuation-dissipation theorem,
\begin{equation}
    D^\text{K}_{0}(\omega)=\coth\left(\frac{\omega}{2T}\right)\left[D_{0}^\text{R}(\omega)-D_{0}^\text{A}(\omega)\right].
\end{equation}

Finally, the interaction part of the action takes the form 
\begin{align}
    S_\text{el-ph}&=- g \int_{-\infty}^\infty \dd t\,\bar{\Psi}_R (\check{\sigma}_0X^{{\rm cl}}+\check{\sigma}_1X^{\rm q})\Psi_R,\label{eq:interaction_action}\\
    &\equiv-\int_{-\infty}^\infty \dd t\,\bar{\Psi}_R(t)\check{V}(t)\Psi_R(t)\nonumber
    \label{Eq:Action_interaction}
\end{align}
with the two-component fermionic field $\Psi_R=(\psi_{1R} \,\,\psi_{2R})^T,$ where $\check{\sigma}_{1/2}$ are Pauli matrices in contour space, and the matrix $\check{V}$ contains the photon fields and coupling strengths in the combined Keldysh and dot basis.

\subsection{Perturbation theory}

The el-ph interaction term  Eq.~\eqref{eq:interaction_action} prevents exact calculation of expectation values and correlation functions. To make progress, we perform a perturbative expansion in the coupling strength, $g$. Performing a Gaussian integration over either the dot electrons or the resonator photons leads to an effective action for the other degree of freedom, which can then be expanded perturbatively to second order in $g$.  

\subsubsection{The photon perspective}

\begin{figure}[t]
    \centering
    \includegraphics[width=\columnwidth]{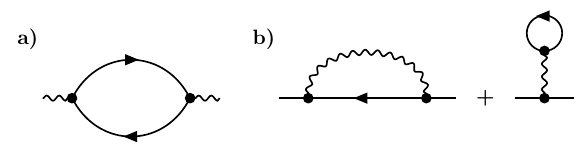}
    \caption{Second-order self-energy diagrams for the photon (a) and the right QD electron (b) (Hartree-Fock diagrams). Wiggly (full) lines represent the photon (electron) propagator and the dots represent the coupling $g$. No further dressing will be considered.}
\label{fig:selfdia}
\end{figure}

The effective action for the photons takes the form
\begin{equation}
    iS_{\text{eff}}[X]=iS_0[X] +\tr\left[\ln(-i(1-\check{G}_0\circ\check{V}))\right],
\end{equation}
where $\circ$ denotes a temporal convolution. 
Expanding the logarithm to first-order leads to a term
\begin{equation}
    S^{(1)}_\text{eff}=2 g n_R \int\dd t\,X^{\rm q} (t),\label{eq:linS}
\end{equation}
where $n_R=-i\int\frac{d\omega}{2\pi}G^<_{RR}(\omega).$
The first-order term can be removed by a constant shift of the classical field, $X^{{\rm cl}}(t)\rightarrow X^{{\rm cl}}(t)+2 g  n_R /\omega_0$. The second-order term introduces a self-energy, identified as the electronic charge susceptibility of the right dot [cf. Fig.~\ref{fig:selfdia}(a)], 
\begin{align}
    \Pi^{{\rm R/A}}&(\omega)=-\frac{i g^2}{2}\int \frac{\dd\omega'}{2\pi}\Big(\G{\rm K}{0}{RR}{\omega'}\G{{\rm A/R}}{0}{RR}{\omega'-\omega}\label{Eq:RetardedPhotonSelfenergy}\\
    &+\G{{\rm R/A}}{0}{RR}{\omega'}\G{\rm K}{0}{RR}{\omega'-\omega}\Big)\nonumber,\\ 
    \Pi^{\rm K}&(\omega)=-\frac{i g^2}{2} \int \frac{\dd\omega'}{2\pi}\Big(\G{{\rm R}}{0}{RR}{\omega'}\G{\rm A}{0}{RR}{\omega'-\omega}\label{Eq:KeldyshPhotonSelfenergy}\\
    &+\G{\rm A}{0}{RR}{\omega'}\G{R}{0}{RR}{\omega'-\omega}\nonumber\\ 
    &+\G{\rm K}{0}{RR}{\omega'}\G{K}{0}{RR}{\omega'-\omega}\Big)\nonumber.
\end{align}
The dressed photon Green function is thereby obtained directly from the Dyson equation as $\check{D}(\omega)=\left[\check{D}_0(\omega)^{-1}-\check{\Pi}(\omega)\right]^{-1}.$

\subsubsection{The electron perspective}

In an analogous perturbative expansion of the electronic part of the action the leading non-vanishing term is of order $g^2$. It yields the Hartree-Fock (HF) self-energies for the right QD [cf. Fig.~\ref{fig:selfdia}(b)], 
\begin{align}
{\Sigma}_{RR}^{\text{R/A}}&(\omega)= 
    - i g^2 \; D_{0}^\text{R/A}(0) \int\frac{\dd\omega'}{2\pi}\;
    G^<_{0,\scaleto{RR}{4pt}}(\omega')
    \label{equ:SigmaRA_fermions}\\
    & + \frac{i g^2}{2}\; \int\frac{\dd\omega'}{2\pi} \;
    \Big(D_{0}^\text{R/A}(\omega-\omega')\G{K}{0}{RR}{\omega'} \nonumber\\ 
    & + D_{0}^\text{K}(\omega-\omega') \G{R/A}{0}{RR}{\omega'}\Big)  ,
    \nonumber\\
{\Sigma}_{RR}^{\text{K}}&(\omega) =
    \frac{ig^2}{2}\; \int\frac{\dd\omega'}{2\pi} 
    \Big(D_{0}^\text{R}(\omega-\omega')\G{R}{0}{RR}{\omega'} 
    \label{equ:SigmaK_fermions}\\
    & + D_{0}^\text{K}(\omega-\omega')\G{K}{0}{RR}{\omega'}\nonumber \\
    &+ D_{0}^\text{A}(\omega-\omega')\G{A}{0}{RR}{\omega'}\Big)\nonumber,
\end{align}
where $G^{<}=\left(G^{\rm K}-G^{\rm R}+G^{\rm A} \right)/2$.
From the corresponding $6\times 6$ Keldysh dot space matrix self-energy, $\check{\Sigma}$, with $\Sigma^\text{R/A/K}_{RR}$ being the only nonzero entries, the dressed dot electron matrix Green function is obtained via the Dyson equation $\check{G}(\omega)=\left[\check{G}_{0}(\omega)^{-1}-\check{\Sigma}(\omega)\right]^{-1}.$ 

Note that our lowest-order perturbative treatment of the self-energies does not include any feedback effects between the resonantor and QD system. These become increasingly important for larger couplings $g$, and could be included through the conserving self-consistent Born approximation~\cite{Frederiksen2007May,Dash2010Mar}, corresponding to a self-consistent dressing of internal lines in the HF diagrams [see Fig.~\ref{fig:selfdia}(b)]. Since this treatment holds the risk of spurious symmetry breaking~\cite{Hewson1993}, we shall not go beyond the undressed HF self-energies in this work.
When excluding feedback effects and the broadening of the bosonic propagators is negligible compared to the fermionic ones ($\Gamma\gg\kappa$), it is consistent to use $\kappa=0$ in $D_0^\text{R/A/K}(\omega)$ when evaluating the electronic self-energies. The resonator simply acts as a probe of photon emission from the voltage-biased QD system. When taking the electronic perspective, we will thus set $\kappa=0$.

In Sec.~\ref{sec:ET} we discuss how perturbative results for expectation values and correlation functions can be computed taking either the photonic or the electronic perspective and employing the corresponding second-order self-energies.  
These results are perturbative in $g$ and exact in all other parameters, such as the bias voltage $V$. In order to assess non-perturbative effects of $g$, we employ the complementary Lindblad master equation approach. 

\subsection{Lindblad master equation}
\label{sec:Lindblad}

The starting point of the Lindblad master equation approach is the equation of motion for the reduced density matrix $\rho(t)$ of a quantum system coupled to reservoirs
\begin{align}
    \partial_t \rho(t) = \mathcal{L}\rho(t).
\label{equ:Lindblad_ME}
\end{align}
In our case the quantum system corresponds to the combined TQD and resonator setup; the electronic leads and the bosonic bath play the role of the reservoirs. 
The action of the Liouville superoperator is defined as
\begin{align}
    \mathcal{L}\rho= -i[\tilde H,\rho]+\sum_\alpha\gamma_\alpha \left( L_\alpha \rho L_\alpha^\dagger-\frac{1}{2}\{L^\dagger_\alpha L_\alpha,\rho\}\right).\label{eq:lbmeq}
\end{align}
The first term describes the coherent evolution of the closed quantum system with Hamiltonian $\tilde H$. Here, $\tilde H$ refers to the Hamiltonian Eqs.~(\ref{eq:Hamiltonian})-(\ref{eq:interaction}) without the fermionic leads and the bosonic bath, but arbitrary $g$. The coupling to the reservoir(s), is included by the Lindblad jump operators in the second term. A pair $\{\gamma_\alpha, L_\alpha\}$ describes the rate and jump operator for a given decay process. Solving Eq.~\eqref{equ:Lindblad_ME} for the steady-state density matrix $\rho_{\rm s}$, steady-state expectation values  of system observables as well as correlation functions can be computed as
\begin{equation}
    \left< A \right> 
    = \Tr\left(A \rho_{\rm s} \right),
\end{equation} 
for an arbitrary QD or resonator operator $A$. 

For quantum systems with several degrees of freedom contained in $\tilde H$, it is often only possible to solve the Lindblad master equation (\ref{equ:Lindblad_ME}) numerically. In our model this is certainly the case and no analytical insights can  be gained using this method. The main numerical challenge is, that for increasing $g/\omega_{0}$ and decreasing $\kappa/\omega_{0}$ a larger number of resonator photons must be kept for convergence. For the couplings $g$ and bosonic dissipation strengths $\kappa$ considered here, however, we typically need to include only 10-20 photons to ensure convergence. The numerical calculations are performed using the QuTiP package~\cite{Johansson2012,Johansson2013}.

The unity partition function obtained from the formal solution of the Lindblad master equation, $Z=\tr[\rho(t)]=1$, may be expressed as a coherent state Keldysh path integral~\cite{Thompson2023Aug, Kamenev2023Jan}. For a purely bosonic or fermionic system the Keldysh action becomes
\begin{align}
\label{equ:mapping_Lind_action}
     &S=\int \dd t\, \left[\bar{\phi}_+i\partial_t\phi_+- \bar{\phi}_-i\partial_t\phi_- 
     -i\mathcal{L}( \bar{\phi}_+, \bar{\phi}_-, \phi_+, \phi_-) \right],\nonumber\\
     &\mathcal{L}( \bar{\phi}_+, \bar{\phi}_-, \phi_+, \phi_-)=-i(H_+-H_-)\\
     &\hspace{2em}+\sum_\alpha \gamma_\alpha\Big[\bar{L}_{-\alpha}L_{+\alpha}\nonumber-\frac{1}{2}(\bar{L}_{+\alpha}L_{+\alpha}+\bar{L}_{-\alpha}L_{-\alpha}) \Big],
\end{align}
with bosonic or fermionic fields $\phi_{\pm}$, where $\pm$ refers to the forward and backward branches of the time contour. The Hamiltonian $H_\pm$ as well as  the jump operators $\bar{L}_{\pm}$ and $L_{\pm}$ are evaluated on the $\pm$ branch respectively. 
 
One of the crucial assumptions in the derivation of the Lindblad
master equation~\eqref{equ:Lindblad_ME} with the right hand side \eqref{eq:lbmeq} is the Markov approximation; all reservoirs coupled to the system must be memoryless~\cite{Breuer2007}. In
our model, we expect this to be the case in the limit of large bias voltage and for a resonator with a high quality factor. 

Under two conditions the action given by the generalization of Eq.~\eqref{equ:mapping_Lind_action} to our coupled QD-resonantor problem becomes identical to that of the Green function formalism  Eq.~\eqref{eq:partition_function}. Firstly the self-energies have to be frequency-independent. This is the case for the limit of infinite bias voltage, where $F_{L/R}(\omega)=\pm 1$, and when treating the resonator-bath interaction in the rotating wave approximation. The latter amounts to evaluating the self-energies at the resonator frequency $\hat{\Sigma}_\text{bath,b}(\omega)\rightarrow \hat{\Sigma}_\text{bath,b}(\omega_0),$ which is reasonable for high-Q resonators with $\kappa \ll \omega_{0}$.
Secondly, we have to take the  following bath-induced rates and jump operators:
\begin{enumerate}
     \item Electron tunneling in from the left lead: $\{2\Gamma,\, c_{L}^\dagger \}$
     \item Electron tunneling out into the right lead: $\{2\Gamma,\, c_{R} \}$
     \item Photon annihilation $\{2\kappa (1+n_{B}(\omega_0)),\, a \}$
     \item Photon creation $\{2\kappa n_{B}(\omega_0),\, a^\dagger \}$
\end{enumerate}
As we are considering low temperatures, $T\ll\kappa\ll\omega_{0}$, the Bose distribution satisfies $n_B(\omega_0)\simeq0$ and the fourth process can safely be omitted. 

Within the Lindblad master equation approach, the Green functions can be computed using the quantum regression theorem (QRT)~\cite{Lax1963,Breuer2007}. Details are given in  Appendix~\ref{app:quantum_regression}. 

\subsection{Applicability of methods}

\begin{figure*}[t]
    \centering
    \includegraphics[width=0.49\linewidth]{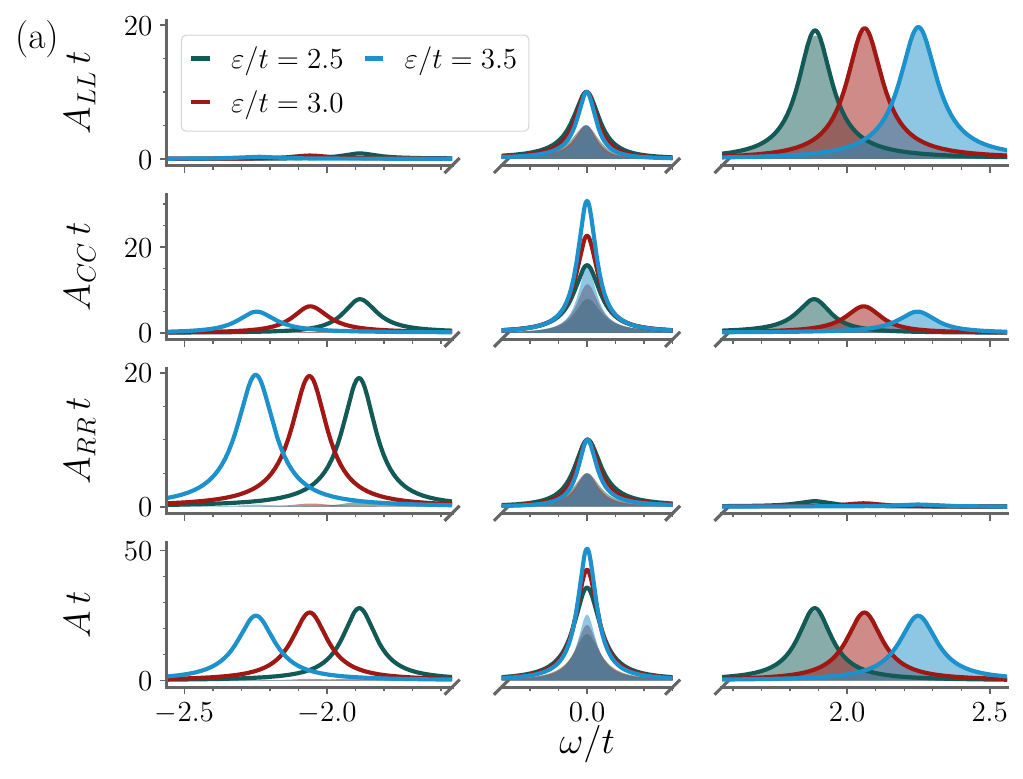}\quad
    \includegraphics[width=0.49\linewidth]{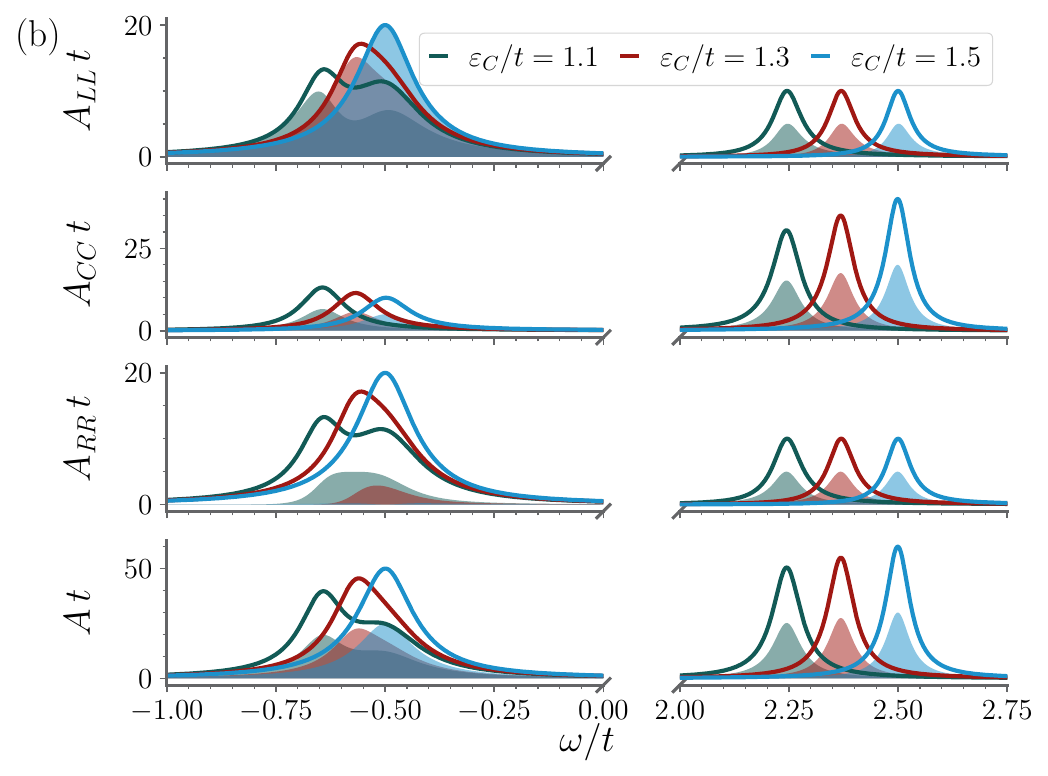}
    \caption{Spectral functions (lines) and corresponding occupational weights (fillings) as a function of frequency for the LTD, $(\text{a})$, and the TTD, $(\text{b})$, configurations. Three different gate settings (green/red/blue) are included for each configuration. The total QD system is in all cases half-filled $N=1.5$. Parts of the frequency axis have been removed for better resolution.}
    \label{fig:spectral_filling}
\end{figure*}
 
In the limit of small el-ph coupling $g$, for large bias voltages $V$ and for small resonator loss $\kappa$ (cf. Table~\ref{tab:my_label}) both approaches are controlled. For parameters within this part of the parameter space, we can simply use the method which allows for a more efficient computation of the quantity of interest. E.g.~the many-body spectra and occupancies of Fig.~\ref{fig:eigenstate_occ_tri} can straightforwardly be obtained from the Lindblad master equation approach, whereas this is much more complicated using Green functions. The frequency dependent transmissions of Fig.~\ref{fig:Transmission_function}, on the other hand, are directly accessible using the Green function formalism. The same holds for the single-particle spectral functions of Fig.~\ref{fig:spectral_filling}; see below. We emphasize that in certain limits the Green function approach allows for analytical insights; see below.      

For large bias voltages, we will investigate the regime of applicability of the perturbative Green function approach by increasing $g$ and comparing it to the results obtained from the Lindblad master equation method which remains accurate. We will see that there is a difference in the range of validity of perturbation theory for respectively electronic, and bosonic observables. Complementary to this, we also study the regime of moderate bias voltages but $g=0$. In this case, the Markovianity of the electronic leads is lost and the Lindblad approach becomes less tenable while the Green function method is exact. This is explored in Section~\ref{sec:Coulomb} and Appendix~\ref{app:PERLind_comparision}, where we introduce the PERLind method~\cite{Kirsanskas2018}. We will see that for static quantities (expectation values) the predictions of the Lindblad approach remain valid, while deviations can be found for dynamical quantities (correlation functions).  

\section{Energy transfer}
\label{sec:ET}

In order to understand the mechanisms underlying the el-ph energy transfer, we first study the electronic spectral weights and the corresponding non-equilibrium populations of the QD system decoupled from the resonator ($g=0$). From these, we calculate the corresponding charge, and energy current across the TQD. Switching on the coupling to the resonator ($g>0$), we study how these quantities change and infer the transferred power. At the same time, we characterize the resonator in terms of photon number and photon spectral function, together with the gain that would be measured in microwave transmission spectroscopy. 

\subsection{Electron populations for $g=0$}
\label{sec:Population_Inversion}

Due to the large bias voltage, all states lie in the bias window, and the nonequilibrium QD electron distribution may be adjusted by tuning the parameters of the TQD system. We define the QD-resolved spectral function as \cite{Kamenev2023Jan}
\begin{align}
    A_{ii}(\omega)= -2\text{Im}\,G^R_{ii}(\omega),
\end{align}
and the local occupational weights
\begin{align}
    n_i(\omega)=-iG^{<}_{ii}(\omega).
\label{Eq:FrequencyOccupationElectrons}
\end{align}
Based on the frequency dependence of these two functions, one may speak of QD-specific population inversions when the occupational weight is shifted towards the high-frequency end of the spectral function. When this is the case, 
excitation energy is available for specific electronic transitions enabling photon emission. This point will be substantiated below in Sec.~\ref{sec:analytlas}, where we analyze the photon self-energy in terms of these quantities.
In the non-interacting case $g=0$, the QD-resolved spectral functions and occupational weights are plotted for the LTD and TTD configurations in Fig.~\ref{fig:spectral_filling} together with the total spectral function $A(\omega)=\sum_i A_{ii}(\omega)$ and total occupational weight $n(\omega)=\sum_i n_i(\omega)$, corresponding to a total filling of $N=\sum_{i}n_{i}$ with $n_{i}=\int\!\frac{d\omega}{2\pi}n_{i}(\omega)$ electrons on the $i$'th quantum dot. For all cases considered here the TQD is at half filling ($N=1.5$).

As seen from the bottom panels in Fig.~\ref{fig:spectral_filling}, considering only the global spectral function would lead to the erroneous conclusion that only the LTD allows for significant population inversion. From a local perspective, which also turns out to be the relevant one when coupling the resonator to a single dot, the TTD is seen to have a nearly perfect population inversion on the right QD when tuning to the degeneracy point at $\varepsilon_C=1.5t$ (blue curves), in the sense that the state of lowest energy is almost completely depleted. 
The emergence of this population inversion is further examined in Appendix~\ref{app:interference}. 
In contrast to this, for the left QD the high-energy state exhibits a smaller occupation than the low-energy state. If the resonator was coupled to the left QD rather than the right one, energy transfer to the resonator would thus not be possible. For a resonator with a finite occupation (e.g. due to a non-zero temperature of the bosonic bath), one would see photon absorption into the electron system rather than photon emission.

As discussed in Sec.~\ref{sec:Model}, this local population inversion is related to a node in the transmission function, $T(\omega)=|G^\text{R}_{RL}(\omega)|^2$. In the large-bias limit, the lesser Green function takes on a particularly simple form, such that the occupational weight becomes
\begin{align}
\label{equ:filling_transmission}
    n_{R}(\omega)&=\Gamma T(\omega),
\end{align}
where the transmission was shown in Fig.~\ref{fig:Transmission_function}. While $n_{R}(\omega)$, (blue shaded area in Fig.~\ref{fig:spectral_filling}) is generally suppressed near the lower single-particle eigenenergy at the degeneracy point $\varepsilon_C=1.5t$, it is only strictly zero at $\omega=\varepsilon_C-t^2/s=-0.5t$. 

This single-particle Green function perspective on population inversion is complementary to the many-body perspective taken in Fig.~\ref{fig:eigenstate_occ_tri}. There, we showed the many-body eigenenergies of the decoupled TQD system ($g=0$, $\Gamma=0$) together with their respective occupation probabilities calculated by the Lindblad master equation for the open system  ($g=0$, $\Gamma>0$). This atomic-limit perspective reveals no local information, whereas the Green function perspective reveals no information about the two- and three-electron spectra. Importantly, the local spectral functions and occupational weights can be extracted from the Lindblad master equation using the quantum regression theorem, as we demonstrate in Appendix~\ref{app:quantum_regression}. Figure~\ref{fig:QRT_spectral_filling} of Appendix~\ref{app:quantum_regression} exemplifies that this yields the same result for the Green function. It also allows us to include the effects of any finite coupling $g>0$ to the resonator, where the Green function results are seen to be slightly off for a coupling of $g=0.1t$.

\subsection{Electric current}
\label{sec:Thermoelectric_current}

The electron current across the TQD system is found as $I^{N}=(I^{N}_{L}-I^{N}_{R})/2$, with local lead-QD currents obtained via the Green functions as~\cite{Haug2008}
\begin{align}
    I_{i}^N= -\Gamma\int\frac{\dd\omega}{2\pi}\left[ \text{Im}\,G_{ii}^\text{K}(\omega)+F^{(0)}_{i}(\omega)A_{ii}(\omega) \right].
\label{equ:particle_current}
\end{align}
Alternatively, the current can be calculated using the steady-state density matrix, obtained by solving Eq.~\eqref{equ:Lindblad_ME} numerically (cf. Appendix~\ref{app:energy_current_Lind})
\begin{align}
\label{equ:current_Lindblad}
    I_L^N=2\Gamma \left( 1-\langle d_L^\dagger d_L \rangle \right),
    \hspace{2em} 
    I_R^N = -2\Gamma \langle d_R^\dagger d_R \rangle. 
\end{align}
For $g=0$, this gives the current which was shown in Fig.~\ref{fig:Transmission_function} to exhibit a pronounced dependence on the detuning $\varepsilon$ for the LTD configuration and on $\varepsilon_C$ for the TTD.

\begin{figure}[t]
    \centering
    \includegraphics[width=1\linewidth]{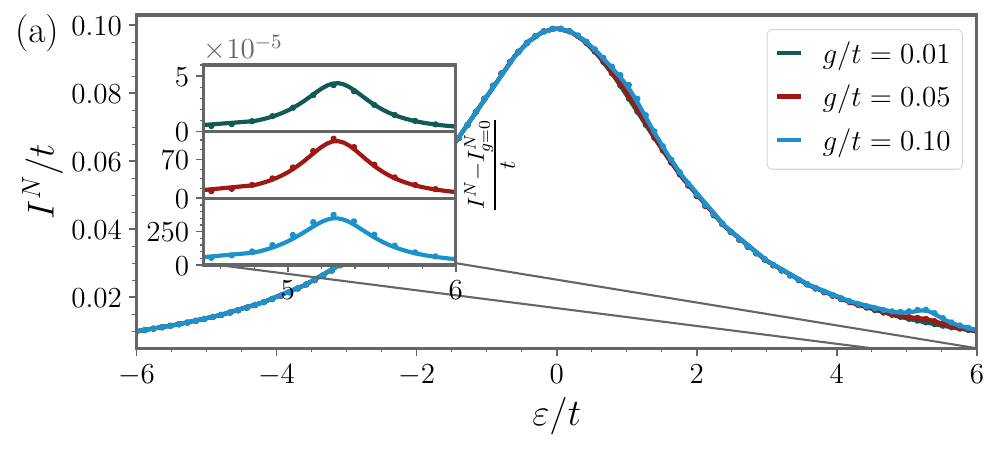}
    \includegraphics[width=1\linewidth]{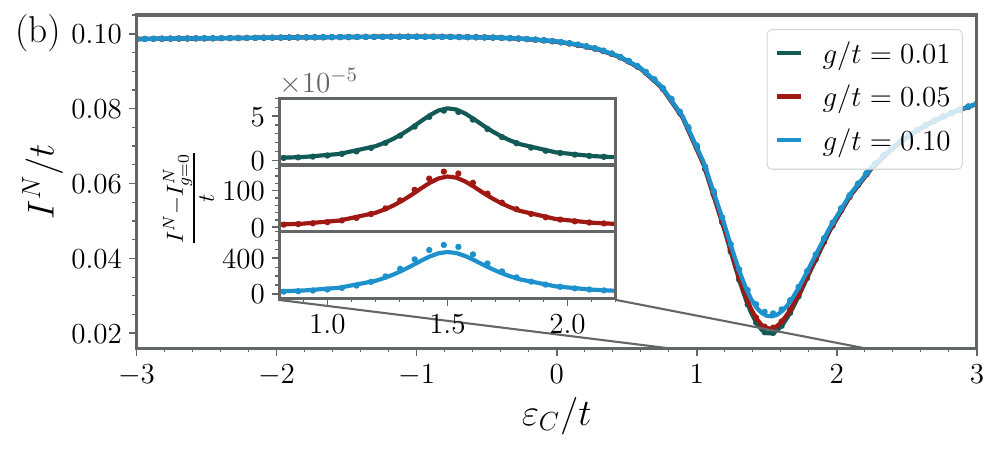}
    \caption{Particle current obtained with the master equation (dots) and the dressed Green function (solid lines) for different couplings $g$. The inset shows the difference between the current through the system decoupled from the resonator and the interacting one. $\text{(a)}$: LTD configuration, $\text{(b)}$: TTD configuration. }
    \label{fig:particle_current}
\end{figure}

In Fig.~\ref{fig:particle_current} we show the corresponding plots in the presence of a finite coupling $g>0$ to the resonator, calculated with both the Green functions (full lines) and the Lindblad master equation (dots). We find good correspondence between the two methods for the couplings considered here: $g/\omega_{0}\approx 0.003-0.03$ and $\kappa/\omega_{0}=0.0017$ (using values from Table~\ref{tab:my_label}), corresponding to a Q-factor of $Q=f/\kappa\sim 100$.  
For nonzero coupling, the current is increased due to photon-assisted inelastic tunneling when the resonator frequency is in resonance with an electronic transition energy, as found also in Ref.~\cite{Jin2011Jul} for a DQD system. However, in Ref.~\cite{Jin2011Jul}, the peak is much sharper resulting from a $\kappa$ taken two orders of magnitude smaller than the value used here. As observed in the insets of Fig.~\ref{fig:particle_current}, the Green function result tends to slightly underestimate this photon-induced increase in current for the larger values of $g$. The agreement between the two methods also strongly depends on $\kappa$, as we will see below.

\begin{figure}[t]
    \centering
    \includegraphics[width=\linewidth]{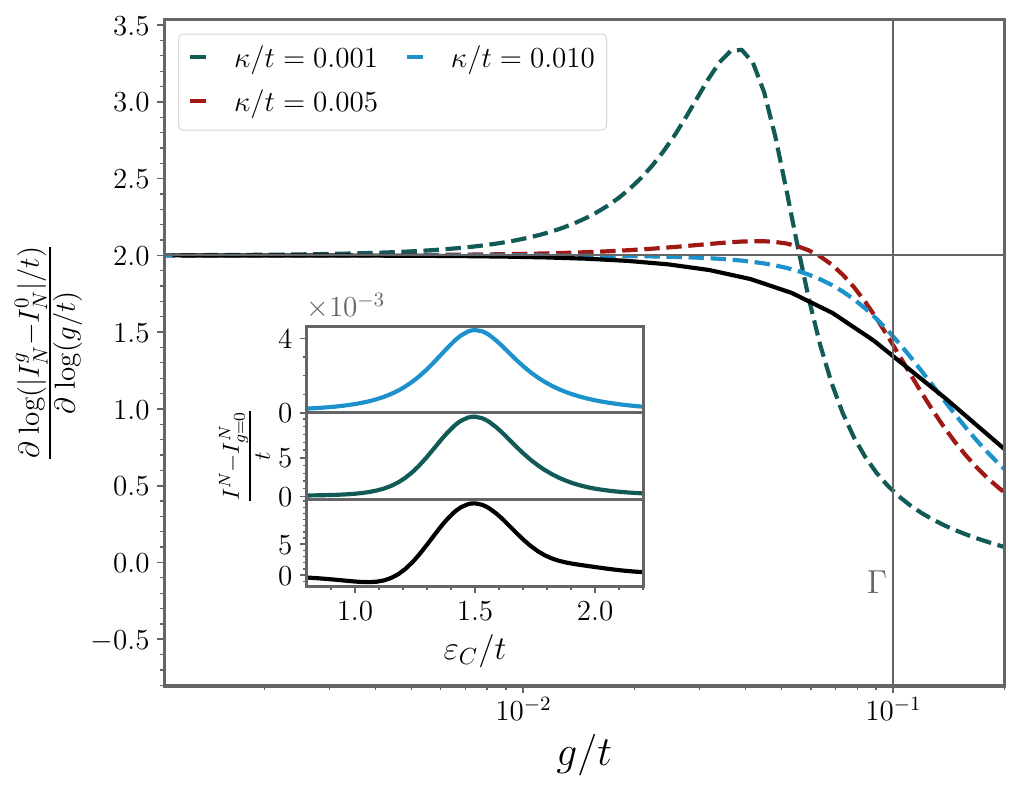}
    \caption{Scaling analysis of the photon-assisted current for the TTD at $\varepsilon_C=1.5t$. We show the logarithmic derivative with respect to $g$ of the difference in current at finite, and zero $g$ against $g$. Full, and dashed curves show the result of the perturbative Green functions (with $\kappa=0$) and Lindblad master equation for different values of $\kappa$ respectively. The vertical line marks the tunneling rate $\Gamma$ and the horizontal line indicates the $g^2$ scaling. The inset shows the change in particle current calculated via the master equation in the vicinity of the degeneracy point $\varepsilon_{C}=1.5t$ with $g=0.1t$ for $\kappa/t=0.01,0.001,0.0005$ from top to bottom, similar to the inset in Fig.~\ref{fig:particle_current}.}
    \label{fig:scaling_behavior}
\end{figure} 
To assess the range of validity of the HF-approximation, we display in Fig.~\ref{fig:scaling_behavior} the $g$-dependence of the electron current enhancement and compare it to the results obtained with the Lindblad master equation (exact in $g$). 
The perturbative $g^2$-scaling of the current enhancement is clearly observed for the smallest values of $g$, and the Green function result starts deviating from this simple behavior near $g\sim\Gamma$. The deviation of the perturbative result from the $g^2-$scaling is due to the fact that when the dressed Green function is calculated using the Dyson equation, terms of higher-order in the expansion parameter are generated. The Lindblad results are seen to depend strongly on the intrinsic decay rate, $\kappa$, of the resonator. 
As demonstrated in Appendix~\ref{app:energy_current_Lind}, this is because the Lindblad master equation incorporates feedback from the increasing photon occupation in the resonator and describes a steady state between the energy pumping from the electrons to the resonator and the intrinsic dissipation in the resonator set by $\kappa$. This important physics is clearly missing within the HF-approximation and therefore the two results do not generally correspond for electron observables like the particle current. 
It is important to note that there is no straightforward limit for $\kappa$ where the results consistently agree.
As $\kappa$ is decreased, the photon number obtained from the Lindblad master equation increases, feedback becomes increasingly important and the agreement with perturbation theory worsens. There is no simple limit in the Lindblad calculation which reproduces the assumption of a zero resonator linewidth along with zero resonator occupation as used in perturbation theory.   

\subsection{Energy current}
\label{sec:energy_current}
As a measure of the energy transfer from the voltage-biased electron system to the resonator, one may calculate the difference between the energy currents into the left dot and out of the right one:
\begin{align}
\Delta I^E_\text{el}
    = -\Gamma\int\frac{\dd\omega}{2\pi}\,\omega \sum_{i=L,R}\left[ \text{Im}\,G_{ii}^K(\omega)+F^{(0)}_{i}(\omega)A_{ii}(\omega) \right].
    \label{eq:energy_dissipation}
\end{align}
Alternatively, using the Lindblad master equation to calculate the steady-state density matrix, the total energy current into the resonator may be calculated as (cf. Appendix~\ref{app:energy_current_Lind}), $I^E_\text{ph}=I^E_{\text{el}\to\text{ph}} + I^E_{\text{ph}\to\text{bath}}$, where 
\begin{align}
    I^E_{\text{el}\to\text{ph}}&=-i\omega_0g\left\langle d^{\dagger}_{R}d^{}_{R}(a^\dagger-a)\right\rangle\label{eq:energy_current_Lind1}\\
    I^E_{\text{ph}\to\text{bath}}&=-2\kappa\omega_0 \left\langle a^\dagger a \right\rangle.\label{eq:energy_current_Lind2}
\end{align}
The first term represents the energy injected into the resonator from the electronic system, including both loss and gain. 
The second term accounts for the energy lost to the bosonic bath. In steady state, the two contributions balance out and the energy of the resonator remains constant. This balancing mechanism is absent in the HF self-energy employed in Eq.~\eqref{eq:energy_dissipation}, and for large el-ph coupling the Lindblad master equation and HF Green function results will therefore differ.

\begin{figure}[t]
    \centering
    \includegraphics[width=0.49\linewidth]{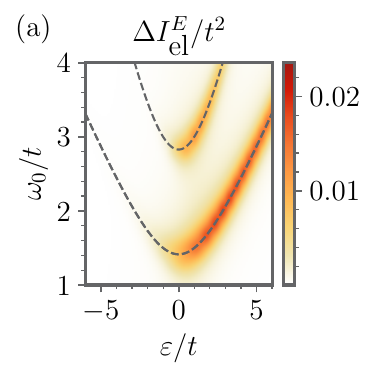}
    \includegraphics[width=0.49\linewidth]{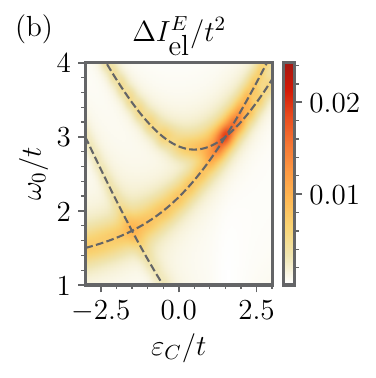}\\
    \includegraphics[width=0.49\linewidth]{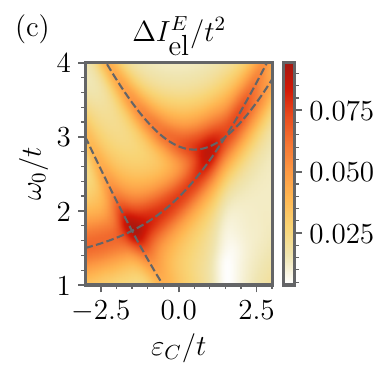}
    \includegraphics[width=0.49\linewidth]{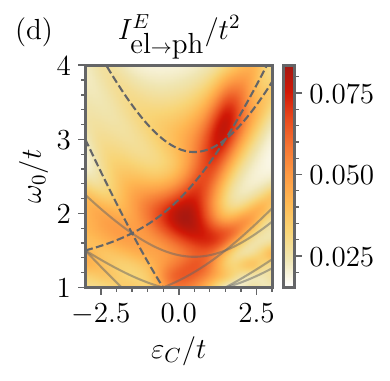}
    \caption{Rate of energy transfer from the TQD electron system into the resonator as a function of the resonator frequency $\omega_0$ and the detunings $\varepsilon_C$ or $\varepsilon$, for the LTD, $(\text{a})$, and the TTD, $(\text{b})$-$(\text{d})$, respectively. The el-ph couplings are $g=0.1t$ ($(\text{a})$, $(\text{b})$) and $g=0.6t$ ($(\text{c})$, $(\text{d})$). Dotted lines indicate the excitation energies $\Delta E$. Upper, and lower left panels are calculated from Eq.~\eqref{eq:energy_dissipation}. $(\text{d})$ is calculated via the Lindblad master equation and shows a marked difference to the HF-approximated result, $(\text{c})$, at the same large coupling. The light gray lines in (d) show the first and second side band excitation energies $\Delta E/2$ and $\Delta E/3$, corresponding to two- and three-photon emission processes.}
    \label{fig:energy_dissipation_grid}
\end{figure}

The energy dissipation is shown in Fig.~\ref{fig:energy_dissipation_grid} for the LTD and the TTD configurations. As expected, this rate of energy transfer is sharply peaked at resonator frequencies which match the energy of a population inverted excitation in the bare TQD electron system (dashed lines).
In contrast to the LTD configuration, which exhibits an extended region of maximal energy transfer, the TTD configuration shows a pronounced maximum at the degeneracy point, $\varepsilon_{C}=1.5 t$. Both observations reflect the population inversions displayed in Figs.~\ref{fig:eigenstate_occ_tri} and~\ref{fig:spectral_filling}. We note, that the width of the energy transfer maxima is set by the electronic tunnel broadening, $\Gamma$.
For the coupling strength, $g=0.1t$, considered in the top panels of Fig.~\ref{fig:energy_dissipation_grid}, the perturbative results agree with the Lindblad master equation (shown in Fig.~\ref{fig:energy_curren_Lind} of Appendix~\ref{app:energy_current_Lind}). As seen from the lower panels, however, choosing the coupling to be $g=0.6t$, perturbation theory fails to agree with the exact master equation results, which now shows signatures of 2- and 3-photon processes.
We expect that the additional regions of large energy transfer, which are absent in the perturbative result, are due to higher-order, and feedback effects. The sidebands, depicted as light gray lines in Fig.~\ref{fig:energy_dissipation_grid} (d), coincide relatively well with the regions of non-zero energy transfer, albeit with some additional shift, which could be due to a renormalization of the resonator frequency.

\begin{figure}[t]
    \centering
    \includegraphics[width=0.49\linewidth]{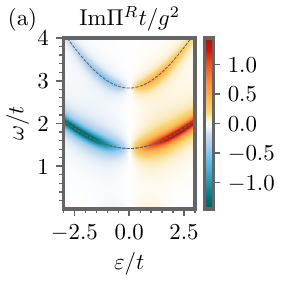}
    \includegraphics[width=0.49\linewidth]{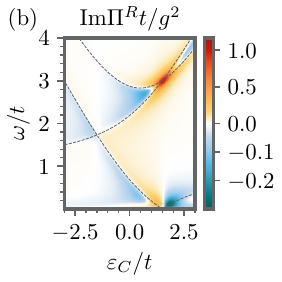}
    \caption{Imaginary part of the photon self-energy as a function of level detuning and frequency. Regions with $\Im \Pi^\text{R}>0$ correspond to photon line-narrowing and a net pumping of the resonator. (a)/(b) show the LTD/TTD configuration results.}
    \label{fig:Charge_susceptibility}
\end{figure}

\subsection{Resonator gain}

\begin{figure}[t]
    \centering
    \includegraphics[width=0.9\linewidth]{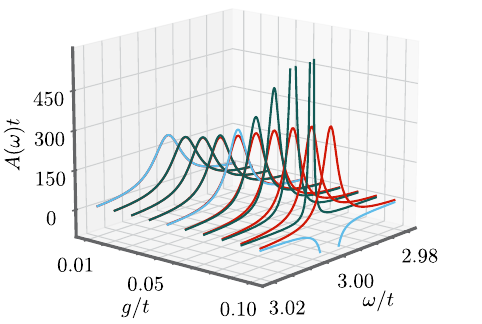}\\
    \caption{Photon spectral functions for the TTD configuration at $\varepsilon_C=1.5t.$ QRT (red lines) and HF-approximated Green functions (green/blue lines) for different values of $g$. Blue curves correspond to the values of $g$ used in Fig.~\ref{fig:transmission_function}.}
    \label{fig:photon_spectral_function}
\end{figure}

From a resonator perspective, the energy transfer from the electron system will increase the photon number in a balance against the losses to the bosonic bath and back to the electrons. This steady-state balance between pumping and losses is manifest in the perturbative retarded photon Green function, from which the lifetime broadening may be obtained as
\begin{equation}
    \text{Im}\left[D^\text{R}(\omega)\right]^{-1}=\frac{2\omega}{\omega_0}\kappa-\Im\Pi^\text{R}(\omega)\equiv \frac{2}{\omega_0} \omega\tilde{\kappa}(\omega).
\end{equation}
Here, we defined an effective (frequency-dependent) damping rate $\tilde{\kappa}$. At zero bias,  $\Im \Pi^R(\omega)$ is always negative for $\omega>0,$ and the el-ph coupling merely increases the dissipation in the photon mode to $\tilde{\kappa}>\kappa$. However, as the voltage bias becomes larger than the electronic excitation energies, the imaginary part of the photon self-energy, i.e. the charge susceptibility of the right QD, changes sign and $\tilde{\kappa}<\kappa.$ This indicates effective pumping and a line-narrowing of the resonator mode due to the biased electron system acting as a gain medium. The imaginary part of the frequency-dependent charge susceptibility is plotted against the respective level detunings, $\varepsilon$ and $\varepsilon_C$, for the LTD and TTD configurations in Fig.~\ref{fig:Charge_susceptibility}. The regions of net pumping are seen to correspond to the regions of largest energy transfer in Fig.~\ref{fig:energy_dissipation_grid} (top panels).

If the loss from the resonator to the bosonic bath is not sufficient to balance the energy transfer from the electrons, $\tilde{\kappa}$ may reach zero and turn negative, signifying a breakdown of perturbation theory as the retarded photon Green function becomes advanced and the spectral function changes sign (see blue and green lines of Fig.~\ref{fig:photon_spectral_function}).
As observed in Fig.~\ref{fig:photon_spectral_function}, no such instability is present within a Lindblad master equation treatment of the TTD setup (red lines), which includes all orders in the el-ph coupling and the non-linear feedback mentioned above. 

Notably, for the photon correlation functions and expectation values shown here, perturbation theory becomes markedly inaccurate already at $g=0.05t$. In contrast, the electronic properties calculated above, all show good correspondence even for $g=0.1t$.

\begin{figure}[t]
    \centering
    \includegraphics[width=0.9\linewidth]{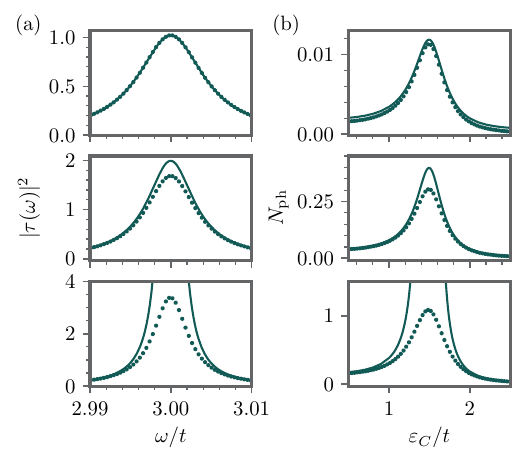}
    \caption{In (a) we show $|\tau(\omega)|^{2}$  versus frequency and in (b) the photon number expectation value versus level detuning. In both cases for the TTD configuration calculated using Lindblad master equation (dots) and HF-approximated Green function (full lines). From top to bottom, $g=(0.01, 0.05, 0.1)t$ corresponding to the blue lines in Fig. \ref{fig:photon_spectral_function}.}
    \label{fig:transmission_function}
\end{figure}

One experimental signature of energy transfer is the gain, defined as the ratio of output, to input power when probing the resonator by a transmission line~\cite{Liu2015Jan}. From the photonic transmission function, obtained as~\cite{Agarwalla2016Sep, Schiro2014May}
\begin{equation}
\tau(\omega)=2i\kappa D^\text{R}(\omega),
\end{equation}
where a value of $|\tau(\omega)|^{2}>1$ signifies gain. $|\tau(\omega)|^{2}$ is plotted for the TTD configuration in Fig.~\ref{fig:transmission_function} (a). A peak is observed at the resonance frequency, $\omega=(2s^2+t^2)/s=3t$, and for all values of $g$ gain is observed at the resonator frequency. Due to the lack of feedback, the gain is seen to be grossly overestimated by the Green function calculation at $g=0.1t$, for this particular set of parameters (see Table~\ref{tab:my_label}). 

\begin{figure}[t]
    \centering
    \includegraphics[width=0.49\linewidth]{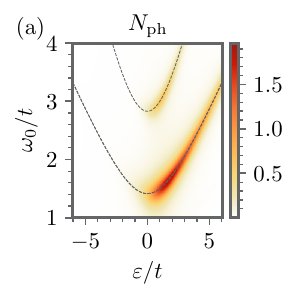}
    \includegraphics[width=0.49\linewidth]{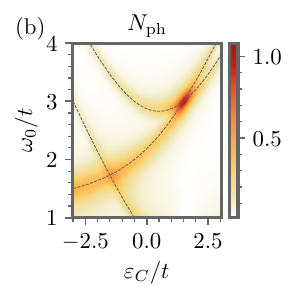}
    \caption{Photon number for the LTD (a) and the TTD (b) configuration as a function of level detuning and resonator frequency  calculated using the Lindbald master equation.}
    \label{fig:Lindblad_photon_number}
\end{figure}

The gain is associated with an increase in photon number, which may be calculated from the Keldysh component of the photon Green function as
\begin{equation}
    1+2N_\text{ph}=\frac{i}{2\omega_0^2}\int\frac{\dd\omega}{2\pi}(\omega^2+\omega_0^2)D^\mathrm{K}(\omega).
\end{equation}
We note that the static shift of $X^{\rm cl}$ performed to eliminate the linear term~\eqref{eq:linS} in the action, must be undone before calculating $D^\text{K}(\omega)$ used in this formula. For the parameters used here, however, this makes a negligible difference. This is plotted in Fig.~\ref{fig:transmission_function} (b), and as for the transmission function, also this photon observable displays a peak at resonance, which is overestimated compared to the result obtained from solving the Lindblad master equation. Fig.~\ref{fig:Lindblad_photon_number} shows the steady-state photon number obtained from solving the Lindblad master equation, which resembles closely the el-ph energy transfer rate shown in Fig.~\ref{fig:energy_dissipation_grid}. The resonator reaches a modest maximum of 1-2 photons when tuned to resonance with the relevant electronic excitation energies (dashed lines) for the LTD and TTD, respectively, where the imaginary part of the self-energy is large and positive (red regions in Fig.~\ref{fig:Charge_susceptibility}) and energy pumping is most efficient (cf. Fig.~\ref{fig:energy_dissipation_grid} (a)/(b)). Decreasing the value of $\kappa$ would lead to an increase in the steady-state photon number.

\subsection{Simple analytical model for lasing}
\label{sec:analytlas}

Approximating the bare spectral function for the electronic system by a sum of Lorentzians centered at the single-particle eigenenergies, it is possible to provide a simple analytical expression for the retarded photonic self-energy Eq.~\eqref{Eq:RetardedPhotonSelfenergy}. This, in turn, allows for an analytical stability analysis, much like it was done in Ref.~\cite{Vyshnevyy2022Feb} for a nanolaser modeled as a system of pumped two-level emitters. To this end, we assume that
\begin{equation}
    G^\text{R}_{RR}(\omega)=\sum_i \frac{Z_{i}}{\omega-E_i+i\gamma_i},
\end{equation}
where $E_i$ are the single-particle eigenenergies, $\gamma_i$ an effective broadening due to the coupling to both leads and $Z_{i}$ is the wave-function renormalization factor or quasiparticle residue. This implies a spectral function given as a weighted sum of Lorentzian distributions
\begin{equation}
    A_{RR}(\omega)=\sum_i \frac{2Z_{i}\gamma_i}{(\omega-E_i)^2+\gamma_i^2},
    \label{Eq:SpectralFunLorentzian}
\end{equation}
with $\int\frac{\dd\omega}{2\pi}A_{RR}(\omega)=1$.
As observed from Fig.~\ref{fig:spectral_filling}, this is always a good approximation in the linear configuration, whereas for the triangular dot it is the case away from the degeneracy point (three peaks) and very close to the degeneracy point (two peaks), but not in the region around the degeneracy point where interference effects change the shape of the two nearly degenerate peaks. Approximating further the nonequilibrium occupational weight of the right QD, $n_{R}(\omega)$, by a scaling factor, $n_{R,i}$, for each spectral peak, such that
\begin{equation}
    G_{RR}^<(\omega)\simeq i \sum_i \frac{2Z_{i}\gamma_i}{(\omega-E_i)^2+\gamma_i^2}n_{R,i},
\end{equation}
the integration in Eq.~\eqref{Eq:RetardedPhotonSelfenergy} can be carried out to give
\begin{align}
    \Pi^\text{R}(\omega)\simeq &\, 2g^2\sum_{ij} n_{R,i}Z_{i}Z_{j}\Bigg(\frac{1}{\omega+(E_i-E_j)+i(\gamma_i+\gamma_j)}\nonumber\\
    &-\frac{1}{\omega-(E_i-E_j)+i(\gamma_i+\gamma_j)}\Bigg).
\end{align}
As we are mainly interested in frequencies close to the resonator frequency, $\omega_{0}$, it is sufficient to consider the term which has poles for $\Re\omega>0,$ allowing us to write
\begin{equation}
    \Pi^\text{R}(\omega)\simeq -2g^2\!\sum_{E_i>E_j}\!\frac{Z_{i}Z_{j}(n_{R,i}-n_{R,j})}{\omega-|E_i-E_j|+i(\gamma_i+\gamma_j)}.
\end{equation}
This shows that the sign and magnitude of the photon self-energy depend on the local population inversion $n_{R,i}-n_{R,j}$ on the right QD, which is then ultimately what determines the efficiency of the energy transfer from the electron system to the resonator, as stated earlier.

Considering the TTD configuration at $\varepsilon_{C}=1.5 t$, where the single-particle spectrum (cf. Fig.~\ref{fig:eigenenergies}) displays a single excited state at energy $E_{3}$ above two perfectly emptied degenerate states at energy $E_{1,2}$, we may approximate the photon self-energy further. Using $\gamma_{1}=\gamma_{2}$ and $Z_{1}=Z_{2}$, it reads
\begin{equation}
    \Pi^\text{R}(\omega>0)\simeq -\frac{4g^2 Z_{1} Z_{3} n_{R,3}}{\omega-\omega_{0}-\Delta+i\gamma_e},
\end{equation}
in terms of a detuning, $\Delta=E_{3}-E_{1}-\omega_{0}$ and with $\gamma_{e}=\gamma_{3}+\gamma_{1}\simeq\alpha\Gamma$, where $\alpha$ is a dimensionless constant of order one. Using this self-energy, one of the poles of the retarded photon Green function is found to cross the real axis at $\omega=\omega_{0}+\kappa\Delta/(\kappa+\gamma_{e})$ when the population inversion reaches the following critical value
\begin{align}
    n_{R,3}^{\ast}=\frac{\kappa\gamma_{e}}{2 g^{2}Z_{1} Z_{3}}\left[1+\left(\frac{\Delta}{\kappa+\gamma_{e}}\right)^{2}\right].\label{eq:criterion}
\end{align}
This criterion for the breakdown of perturbation theory, where the retarded photon Green function becomes advanced, corresponds to the lasing instability criterion obtained from the Maxwell-Bloch equations~\cite{Nazarov2013Jan}, albeit without the usual macroscopic number of atoms in the gain medium to ensure a population inversion much larger than one. 
\begin{figure}[t]
    \centering
    \includegraphics[width=0.95\linewidth]{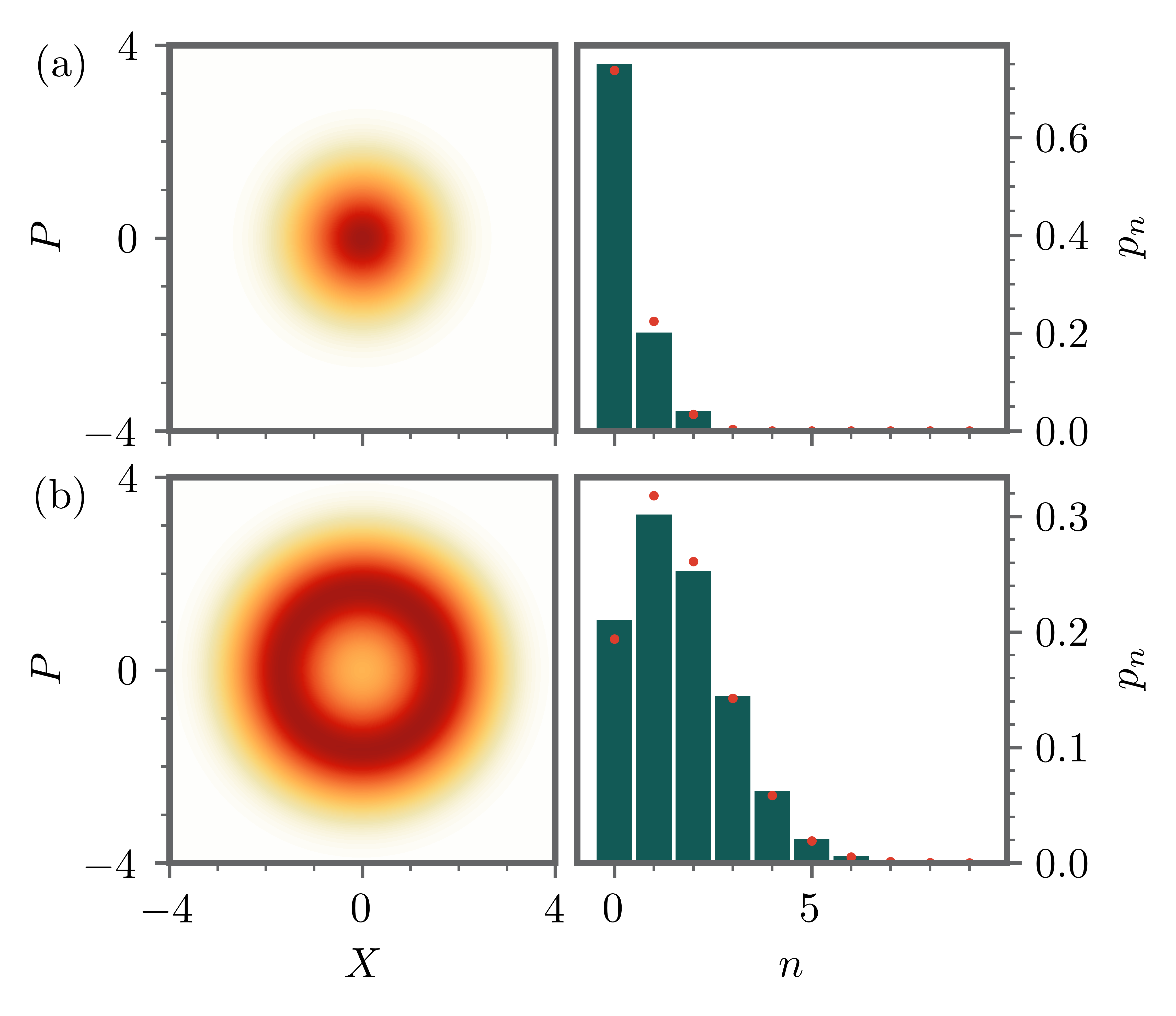}
    \caption{Wigner function (left) and photon number probability distribution (right) for the TTD at resonance $\varepsilon_C=1.5t$. In (a)/(b) the coupling is $g=0.05t/0.15t$. Red dots correspond to a coherent state with the expectation value of $N_\text{ph}$ obtained using $\rho_{\rm{s}}$, which is $\langle N_\text{ph} \rangle=0.3$ in (a) and $\langle N_\text{ph} \rangle=1.64$ in (b). }
    \label{fig:wigner_function}
\end{figure}

For the parameters in Table~\ref{tab:my_label}, a good fit to the actual Green functions is obtained for $\gamma_{1}/t=0.083$, $\gamma_{3}/t=0.033$, (i.e. $\alpha=1.2$), $Z_{1}=0.41$, $Z_{1}=0.17$, and $n_{R,3}=0.50$, leading to a nearly normalized spectral function with $\int\frac{\dd\omega}{2\pi}A_{RR}(\omega)\simeq 0.99$. For these values, one finds from Eq.~\eqref{eq:criterion} with $\Delta=0$ that the instability would take place at $g/t\simeq 0.092$, which corresponds well to the coupling strength at which the photon spectral function obtained from the Green function suddenly becomes negative (cf. Fig.~\ref{fig:photon_spectral_function}).

From the steady-state density matrix, one may calculate the corresponding photon Wigner function~\cite{Walls2008}. As shown in Fig.~\ref{fig:wigner_function}, the phase-space portraits and photon number distributions are consistent with the resonator being in a (phase-unlocked) coherent state, where the photon number with the largest probability is zero for $g/t=0.05$ and one for $g/t=0.15$. This resembles the experimental observation for a single, or two DQD systems coupled to a microwave resonator in Ref.~\cite{Liu2015Jan}, except for the difference in the photon number expectation value. They remain of the order of one here since we consider a resonator with a quality factor as low as $Q=f_{0}/\kappa\sim 100$, as compared to the photon distribution of Ref.~\cite{Liu2015Jan} peaking at $N_\text{ph}\simeq 8000$ (for two DQDs) in a resonator with $Q\simeq 3000$.
Whereas an underdamped resonator mode with $Q\sim 100$ is clearly not optimal for masing, it is comparable to typical values for dominant underdamped vibrational modes in biochemistry~\cite{Duan2020May}.

\section{Effects of Coulomb interaction}
\label{sec:Coulomb}

As discussed in the introduction, one of the main advantages of using QD-arrays to simulate hydrocarbon molecules is the native strong Coulomb interaction. For a typical hydrocarbon molecule, the Ohno representation indicates a nearest neighbor Coulomb interaction as large as 60-70\% of the onsite interaction~\cite{Ohno1964Jan}. Coulomb interaction effects will clearly be important, also for the energy transfer processes which we have considered above.

In order to incorporate Coulomb interaction in the Green function approach it is necessary to include additional self-energy and vertex corrections. Here, we circumvent this rather non-trivial step, by using the Lindblad master equation, for which the nearest-neighbor Coulomb interactions pose no additional complications.
    
\subsection{Nearest-neighbor Coulomb interaction}

\begin{figure}[t]
    \centering\includegraphics[width=0.48\linewidth]{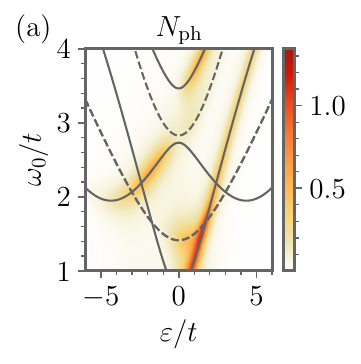}\includegraphics[width=0.48\linewidth]{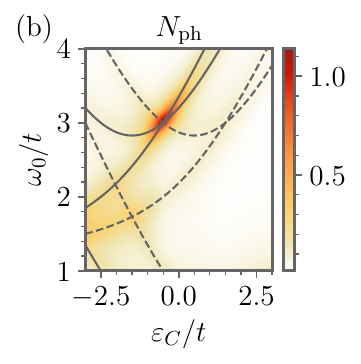}
    \caption{Resonator occupation as a function of $\varepsilon$ or $\varepsilon_C$ and $\omega_0$ and with nearest-neighbor Coulomb interaction $U=20t$ and $\tilde{U}=0.9U$, for the LTD, $(\text{a})$, and TTD, $(\text{b})$, configuration. The dotted lines show the excitation energies of the isolated ($g=0, \Gamma=0$) TQD Hamiltonian in the one-particle sector and the solid lines in the two-particle sector.}
    \label{fig:bosonic_occupation_U}
\end{figure}

Our simplification to spin-polarized electrons has reduced the electronic Hilbert space dimension of the TQD from 64 to 8, which provides a significant numerical advantage when solving the Lindblad master equation for the coupled QD-resonator problem. This is of particular importance for high-$Q$ resonators demanding a large photon Fock space. Within this simplified spin-polarized system, one may still inquire about the effects of nearest-neighbor Coulomb interactions. Here they are included as
\begin{equation}
    H_U
    = U d_L^\dagger d_L  d_C^\dagger d_C 
    + U d_C^\dagger d_C d_R^\dagger d_R 
    + \tilde{U} d_L^\dagger d_L d_R^\dagger d_R,
\end{equation}
which preserves the mirror symmetry of the system. Breaking this symmetry would lift the degeneracy in the two-particle sector, thereby weakening the resonant energy transfer, in particular when the degeneracy is lifted by more than the $\Gamma$ broadening. 
When $\tilde{U}=U$, all two, and three-particle eigenenergies are simply shifted by a factor $U$ and $3U$, respectively and the excitation energies remain unchanged. 
For $\tilde{U}\neq U$, the excitation energies of the one- and two-particle sectors no longer coincide and the degeneracies in the TTD configuration occur at two different values of $\varepsilon_C$.
The largest shift is found for $\tilde{U}=0$, where the degeneracy points for the one, and two-particle sectors are now shifted to $\varepsilon_C=(t^2-s(s\mp U/2))/s$, respectively.  

The resonator occupation together with the excitation energies of the electronic many-body Hamiltonian $H_\text{el}+H_U$ is shown in Fig.~\ref{fig:bosonic_occupation_U} for $\tilde{U} = 0.9 U$ and for both configurations.
$U$ is chosen to be much larger than the other electronic energy scales, except for the bias voltage, rendering the TQD system Coulomb blockaded with energetically well-separated electron number sectors.
As emphasized, the one- and two-particle eigenenergies split up in both cases. For the LTD configuration, this leads to a breaking of the equidistant structure of the eigenenergies, much like the effect of a finite $\varepsilon_{C}$ for $U=\tilde{U}=0$ in (cf. Fig.~\ref{fig:eigenenergies}), and the energy transfer follows only one of the excitations.
In the TTD configuration the energy transfer peak moves with the degeneracy in the two-particle sector. This is consistent with  Fig.~\ref{fig:eigenstate_occ_tri}, showing only a population inversion in the two-particle sector.
Altogether, the nearest-neighbor Coulomb interaction will not degrade the resonant energy-transfer, but mainly shift its maximum in the space of gate-tuning and resonator frequency.

\subsection{Effects of intradot Coulomb interaction and finite bias}

\begin{figure}[t]
    \centering
    \includegraphics[width=0.9\linewidth]{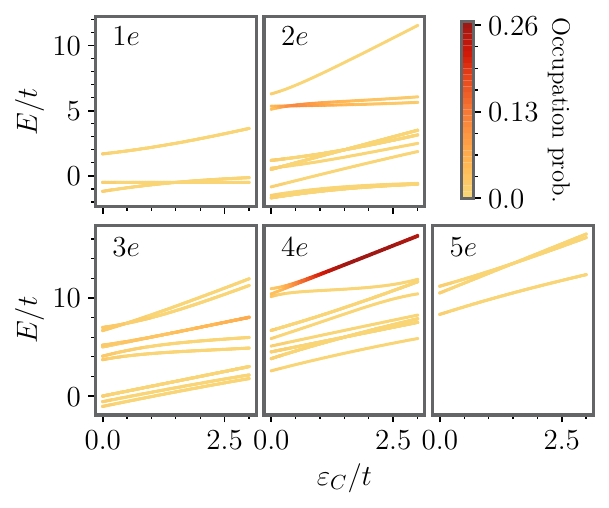}
    \caption{Eigenenergies of the closed system (interacting TTD $U=5t$) and occupations obtained when coupling to the leads at large bias $V=100t$ calculated with PERLind.}
    \label{fig:eigenstate_occupation_interacting_Vinf}
\end{figure}
With available magnetic fields of up to 10 Tesla, our simplification to spin-polarized electrons could be of direct relevance to systems with a $g$-factor of two, when all energy scales are safely below $1\,$meV. With typical QD charging energy of the order of $1\,$meV, say, this is just barely possible. Nevertheless, current high-impedance resonators providing for a strong coupling to the QD-array rely on superconductors, which will be quenched by even weaker magnetic fields. In reality, the question therefore remains, as to how closely our spin-polarized system resembles the real spinful electronic problem. 

In Fig.~\ref{fig:eigenstate_occupation_interacting_Vinf}, we show the full many-body spectrum together with the occupation probabilities of the corresponding 1e-5e eigenstates for the spinful TQD system ($g=0$, $\Gamma>0$) with onsite Coulomb energy, $U=5t$. The bias voltage is now increased to $V=100t$, to retain the infinite-bias limit and thus the validity of the Markov approximation underlying the Lindblad master equation. At this large bias voltage the population inversion of this full system bears little resemblance to the spin-polarized result in Fig.~\ref{fig:eigenstate_occ_tri}, although some of the excited states are still seen to have a large occupation probability, allowing for population inversion.

In order to assess the outcome at smaller bias voltages, we employ the so-called position and energy-resolving Lindblad approach (PERLind) described in Ref.~\cite{Kirsanskas2018}. Here the Lindblad jump operators are expressed in the eigenbasis of the Hamiltonian, allowing for a Fermi function in the jump operators, so as to keep track of filled, and empty states in both the source and drain contacts. Formally, this leads to non-Markovian memory effects on a time scale set by the inverse temperature, but in practice, the currents calculated with this method have been shown to correspond well to exact results for non-interacting problems~\cite{Kirsanskas2018}. In Appendix~\ref{app:PERLind_comparision}, we provide a comparison of the exact Green function results with the finite-bias PERLind calculation in the spin-polarized case. 
There, it is shown that results obtained by this approximate method deviate from exact ones when considering dynamical quantities. 
Nevertheless, static observables such as the occupation are well reproduced, and can safely be calculated using the PERLind approach.
\begin{figure}[t]
    \centering
    \includegraphics[width=0.9\linewidth]{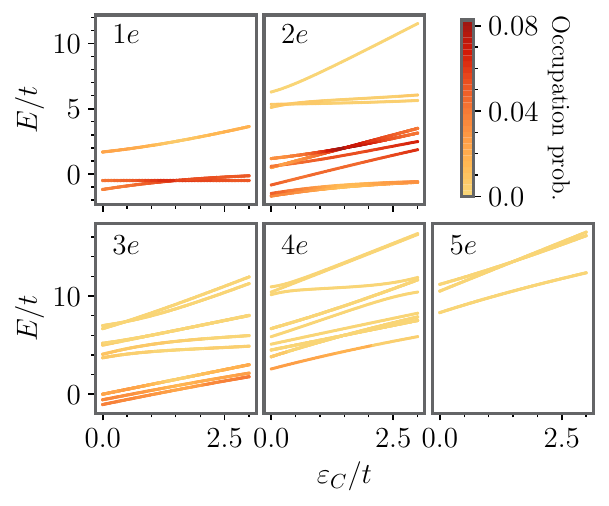}
    \caption{Eigenenergies of the closed system (interacting TTD with $U=5t$) and occupations obtained when coupling to the leads at finite bias $V=8t$ calculated with PERLind.}
    \label{fig:eigenstate_occupation_interacting_Vfin}
\end{figure}
Results for the eigenenergies and the occupation are shown in Fig.~\ref{fig:eigenstate_occupation_interacting_Vfin} for $V=8 t$, which is larger than all electronic energy scales, but not larger than all many-body eigenenergies. The resemblance with the spin-polarized scenario depicted in Fig.~\ref{fig:eigenstate_occ_tri} is now striking, and the dominant population inversions are again found in the two-electron sector.  

\section{Summary and outlook}
\label{sec:discuss}

We have presented the idea of using small arrays of QD-resonator hybrids as analog simulators of hydrocarbon molecules. They operate at energy, frequency, and temperature scales which are scaled down in equal proportions by approximately four orders of magnitude from their relevant values in actual molecules. As an illustration of this idea, we have analyzed a voltage-biased TQD-resonator system as a simulator of current-induced vibrational pumping in single-molecule junctions.

More specifically, we have analyzed the resonant energy transfer from a voltage-biased TQD to a single microwave resonator coupled capacitively to one of the three QDs. This extends earlier studies of maser action from biased DQDs to a system with three molecular orbitals providing for a gate-tunable transmission node, giving rise to a pronounced minimum in the electric current through the TTD, interrelated to a maximum in the rate of energy transfer to the resonator. The presence of these two simultaneous gate-tuned extrema in the respective photon, and electron sectors makes this device particularly suited for a first demonstration of non-trivial molecular simulation capability.

Our analysis is carried out by means of perturbation theory in the Keldysh Green function formalism as well as Lindblad master equations. For weak el-ph coupling and large bias voltage, we obtain good agreement for all calculated observables and correlation functions, and we demonstrate that the corresponding single- and many-particle perspectives offered by either approach provide supplementary insights into the local, and global population inversions governing the resonant energy transfer. Alongside with the analysis of the actual energy transfer, this work therefore also serves the methodological purpose of comparing these two widely used methods in a nontrivial example, which reveals their respective strengths and weaknesses.

We calculate the current-induced energy transfer for bias voltage larger than all other energy scales, and for a resonator with a moderate quality factor, which is merely large enough to justify the rotating-wave approximation.
An actual experiment on a  TQD-resonator device which conforms to these expectations can henceforth be used to simulate the theoretically more challenging regimes with no clear separation of energy scales. In this simulator, the current-induced excitation of vibrational modes, which may lead to vibrational instabilities~\cite{Schulze2008Jun, Diez-Perez2009Nov}, can be monitored as a marked increase in the resonator photon number, controlled mainly by the resonator quality factor. In some respects, the simultaneous access to electron and photon degrees of freedom makes the QD-resonator system a downscaled simulator of Raman enabled single-molecule junctions~\cite{Ward2011Jan, Bi2018Apr, Bi2020Feb}. Given the electrical tunability of the simulator, it should therefore be possible to simulate the different mechanisms for current-induced bond rupture and dissociation, which have been identified recently using hierarchical equations of motion (HEOM)~\cite{Erpenbeck2018Jun, Ke2021Jun}. 

Further simulation perspectives for QD-resonator devices should include the somewhat simpler system of a single QD coupled to a resonator, which realizes the venerable Anderson-Holstein model. With experimental access to both photon and electron properties, it would be highly valuable to see some of the many theoretical results, which have been obtained in different parameter regimes (cf. e.g. Refs.~\cite{Mitra2004Jun, Kaat2005Apr, Koch2006Apr, Batge2021Jun, Rudge2023Mar, Werner2024Feb} and references therein), realized in actual experimental simulations. 

Simulating electron-vibron dynamics in QD-resonator arrays, should also be of relevance to biochemistry, where the resonance of vibrational modes with electronic transition energies has been suggested to play a key role in the remarkably efficient energy transfer taking place across light-harvesting antenna complexes responsible for photosynthesis~\cite{Dijkstra2015Feb, Novoderezhkin2015Nov, Maly2016May}. Efficient absorption of microwave photons in DQDs has recently been demonstrated~\cite{Khan2021Aug}, and the ensuing excitation dynamics in an array of coupled DQD-resonator units could simulate the energy transfer across a rudimentary model of the Fenna-Matthews-Olson (FMO) complex~\cite{Matthews1979Jun, Nalbach2015Feb}. This idea was pursued already in Ref.~\onlinecite{Potocnik2018Jun} simulating energy transfer across three down-scaled chlorophylls represented by transmon qubits. The QD-simulator proposed here, would add an extra microscopic layer by including the electronic degree of freedom so as to simulate the actual polaron formation involved in the energy transfer taking place in the FMO-complex~\cite{Halpin2014Mar, Duan2020May}. Detailed Holstein-like models are available for FMO and their dynamics are currently being studied using Lindblad master equations as well as HEOM~\cite{Delgado2023Jun}. Given the complexity of the problem, analog simulators based on QD-resonator hybrids should be a valuable supplement to this subfield of computational biology.  


\section*{Acknowledgements}

We acknowledge useful discussions with Enrico Arrigoni and Dante Kennes. This work is supported by Novo Nordisk Foundation grant NNF20OC0060019 (CH) and by the Deutsche Forschungsgemeinschaft (DFG,
German Research Foundation) via RTG 1995 (MC). The authors gratefully acknowledge computing time on the supercomputer JURECA~\cite{JURECA} at Forschungszentrum Jülich under grant no. enhancerg. 

\appendix
    
\section{Interference in the spectrum}
\label{app:interference}

To understand the interference occurring in the spectral weight and filling at $g=0$ as discussed in Sec.~\ref{sec:Population_Inversion}, we analyze how the different eigenstates contribute to these quantities for the right QD of the TTD system. 
We concentrate on the case of narrow peaks for small $\Gamma$.
For these analytic considerations, we use the Green function approach.

First, we take a closer look at the spectral function by rotating the $R$-component of the retarded Green function into the eigenspace of the 
inverse-Green function matrix $\left[G^\text{R}(0)\right]^{-1}$ for $g=0$, containing the lead self-energy. This allows us to write
\begin{equation}
    \text{Im}\, G_{RR}^\text{R}(\omega) 
    = \sum_{\alpha=1}^3 \text{Im} \left( S_{R\alpha }\frac{1}{\omega-E_\alpha } \left(S^{-1}\right)_{\alpha R} \right),
\end{equation}
where $\alpha$ runs over the eigenspace with complex eigenenergies $E_\alpha$ and the contributions are weighted by the product of elements of the transformation matrix, $S$, diagonalizing $\left[G^\text{R}(0)\right]^{-1}$.
It can be shown that for small $\Gamma$, the real part of these weight factors gives the largest contribution and is always positive, meaning that the contributions always add up, leading to the increase of the spectral function at the degeneracy point.

Next, we analyze the component of the lesser Green function, giving the occupational weights. Considering the infinite-bias limit, the lesser self-energy obtains a frequency-independent form, leading to
\begin{equation}
    G^<_{RR}(\omega) 
    = \ii\Gamma |G^\text{R}_{RL}(\omega)|^2.
\end{equation}
As stated in Eq.~\eqref{equ:filling_transmission}, the occupational weight of the right dot is therefore proportional to the transmission function $T(\omega)=\left| G_{RL}^\text{R}(\omega) \right|^2$.

The transmission
\begin{align}
    \left|G^\text{R}_{RL}(\omega)\right|^2 
    &= \left| \sum_{\alpha=1}^3 S_{R\alpha}\frac{1}{\omega-E_\alpha} \left(S^{-1}\right)_{\alpha L} \right|^2
\end{align}
can become small when the addends cancel each other.
Again the imaginary contributions to the weight factors $S_{R\alpha}\left(S^{-1}\right)_{\alpha L}$ are negligible compared to the real parts. 
The real parts for the two degenerate eigenenergies can be shown to be finite and have different signs. 
This means that the two contributions are subtracted from each other and the transmission is decreased, leading to the population inversion, we see in the spectrum of the right dot.

\section{Quantum regression theorem}
\label{app:quantum_regression}

To compare the spectral properties calculated from the Green function with the results of the master equation, one can utilize the QRT~\cite{Lax1963,Breuer2007}, which allows the calculation of two-time correlation functions for $t\geq 0$~\cite{Wang2022}. 
Using the time-translational invariance in steady state, they take the form
\begin{equation}
\begin{aligned}
    \left\langle A(t)B \right\rangle &= \Tr\left( A e^{\mathcal{L}t} B\rho_{\rm{s}} \right), \\
    \left\langle BA(t) \right\rangle &= \Tr\left( A e^{\mathcal{L}t} \rho_{\rm{s}} B \right).
    \label{eq:QRT}
\end{aligned}
\end{equation}
An additional minus sign in the first term of the Liouvillian entering Eq.~\eqref{eq:QRT} has to be included, when the jump operators as well as $A$ are fermionic~\cite{Schwarz2016}.
Following the calculations in~\cite{Scarlatella2019} and utilizing~\cite{Dorda2013,Schwarz2016}, one can express correlation functions in the basis of left and right eigenstates of the Liouvillian  
\begin{equation}
    \mathcal{L}r_\nu = \lambda_\nu r_\nu,
    \hspace{1cm}
    \mathcal{L}^\dagger l_\nu = \lambda^*_\nu l_\nu,
\end{equation}
in a way that is reminiscent of the Lehmann representation. The following expression for the retarded Green function is obtained \cite{Scarlatella2019}:
\begin{equation}
    G_{ij}^\text{R}(\omega) = \sum_\nu \frac{\Tr\left(d_i r_\nu\right) \Tr\left(l_\nu^\dagger \left[ d_j^\dagger, \rho_{\rm{s}} \right]_\zeta \right)}{\omega - i\lambda_\nu},
\end{equation}
with $\zeta=+1$ for bosons and $\zeta=-1$ for fermions.
Analogously, an expression for the lesser Green function can be found as
\begin{equation}
\begin{aligned}
    G_{ij}^<(\omega) = \zeta\sum_\nu \Bigg[ 
    &\frac{\Tr\left( d_i r_\nu \right)\Tr\left( l_\nu^\dagger \rho_{\rm{s}} d_j^\dagger \right)}{\omega-i\lambda_\nu} \\
    &- \frac{\Tr\left( d_j^\dagger r_\nu \right)\Tr\left( l_\nu^\dagger d_i \rho_{\rm{s}} \right)}{\omega+i\lambda_\nu}
    \Bigg].
\end{aligned}
\end{equation}

\begin{figure}[t]
    \centering
    \includegraphics[width=\linewidth]{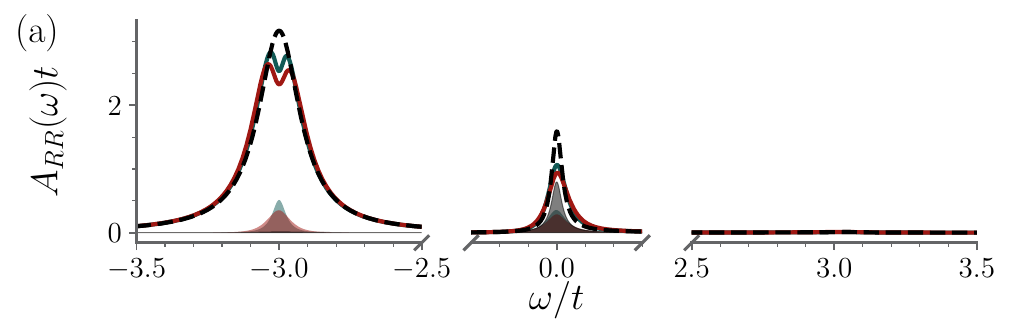}
    \includegraphics[width=\linewidth]{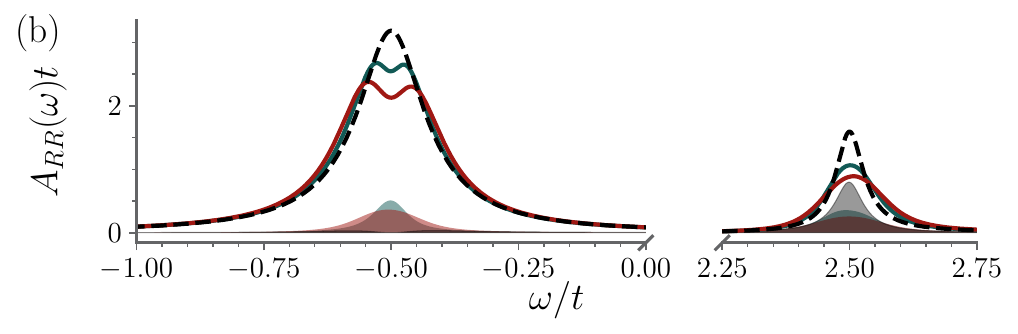}
    \caption{Spectral function (lines) and occupational weight (filled regions) of the right QD in the LTD and TTD configuration for $g=0$ (black dashed) and $g=0.1t$ (red/green full) at resonance with the resonator ($\omega_{0}=3t$). Green functions in the HF-approximation (green) and Lindblad master equations (red) compared to the decoupled case, $g=0$ (black). $(\text{a})$: for the LTD configuration at $\varepsilon =2\sqrt{\omega_0^2-2t^2} =2\sqrt{7}t$, $(\text{b})$: for the TTD setup at $\varepsilon_C=1.5t$.}
    \label{fig:QRT_spectral_filling}
\end{figure}

For the coupled system ($g>0$, $\Gamma>0$) the comparison of the spectral function and filling calculated via perturbation theory and with the QRT are shown in Fig.~\ref{fig:QRT_spectral_filling} for the LTD (a) as well as the TTD (b) configuration on resonance.
In both cases, the coupling of the QD system to the resonator 
leads to the expected decrease in the population inversion with the deexcitation process leading to the observed energy transfer.
Perturbation theory, although showing the same features as the master equation, seems to underestimate the effects of the coupling. This is in line with the observation for the particle current in Fig. \ref{fig:particle_current}.

\section{Currents from the master equation}
\label{app:energy_current_Lind}

The particle current through a specific site $i\in\{ L,C,R \}$ in the TQD system is defined via the change of particle number $I^N_i = \frac{\dd}{\dd t}\langle n_i(t) \rangle$~\cite{Kirsanskas2018}, with $n_i=d_i^\dagger d_i$. By using the master equation \eqref{equ:Lindblad_ME}, this can be evaluated as
\begin{equation}
    I^N_i 
    = \frac{\dd}{\dd t} \Tr\left( n_i \rho(t) \right)
    = \Tr\left( n_i \mathcal{L}\rho(t) \right)
    \rightarrow \Tr\left( n_i \mathcal{L}\rho_{\rm{s}} \right),
\end{equation}
where the steady-state limit was taken in the last step.
To analyze this expression further, the Liouvillian of the system is plugged in.
Because the Hamiltonian without the leads and bosonic bath ($\Gamma=0$, $\kappa=0$) commutes with the particle number operator, the first contribution
\begin{equation}
    -i\Tr\left( n_i \left[H,\rho_{\rm{s}}\right] \right)
    = -i\Tr\left(\left[n_i,H\right]\rho_{\rm{s}} \right) = 0
\end{equation}
vanishes.
Using the invariance under cyclic permutation of the trace, the dissipative contributions can be brought in the form
\begin{equation}
\begin{aligned}
\label{equ:Lindblad_dissipative_current}
    &\sum_\alpha \gamma_\alpha \Tr\left( n_i L_\alpha\rho_{\rm{s}} L_\alpha^\dagger - \frac12 n_i L_\alpha^\dagger L_\alpha \rho_{\rm{s}} - \frac12 n_i \rho_{\rm{s}} L_\alpha^\dagger L_\alpha \right) \\
    = & \sum_j \gamma_j \left( \langle L_\alpha^\dagger n_i L_\alpha \rangle - \frac12 \langle n_i L_\alpha^\dagger L_\alpha \rangle -\frac12 \langle L_\alpha^\dagger L_\alpha n_i \rangle \right).
\end{aligned}
\end{equation}
Because bosonic and fermionic operators commute, the bosonic dissipator does not contribute. 
For $i=L$, we obtain a non-vanishing contribution only from $L_\alpha=c_L^\dagger$
\begin{equation}
    \langle c_L n_L c_L^\dagger \rangle - \frac12 \langle n_L c_L c_L^\dagger \rangle -\frac12 \langle c_L c_L^\dagger n_L \rangle
     = 1-\langle n_L \rangle
\end{equation}
and for $i=R$, only from $L_\alpha=c_R$
\begin{equation}
    \langle c_R^\dagger n_R c_R \rangle - \frac12 \langle n_R c_R^\dagger c_R \rangle -\frac12 \langle c_R^\dagger c_R n_R \rangle
     = -\langle n_R \rangle.
\end{equation}
Which provides the expressions given in Eq.~\eqref{equ:current_Lindblad}.

\begin{figure}[t]
    \centering
    \includegraphics[width=0.48\linewidth]{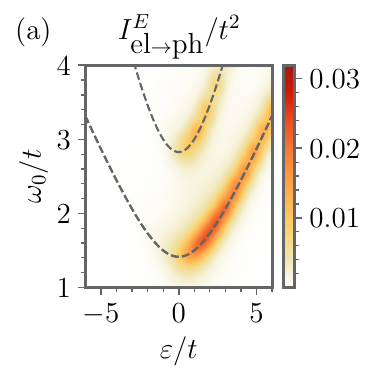}
    \includegraphics[width=0.48\linewidth]{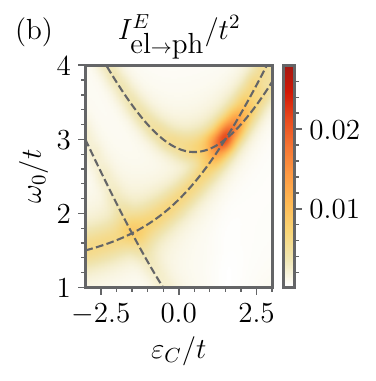}
    \caption{Energy current from the fermionic system into the resonator comparable to Fig. \ref{fig:energy_dissipation_grid} calculated with the Lindblad master equation. $(\text{a})$: for the LTD setup, $(\text{b})$: for the TTD one.}
    \label{fig:energy_curren_Lind}
\end{figure}

Similarly to the approach outlined above, we can calculate the energy current into the resonator by evaluating the expectation value $I^E=\frac{\dd}{\dd t}\left\langle H_\text{ph} \right\rangle$.
Here, we have to consider two non-vanishing contributions:
\begin{equation}
\begin{aligned}
    \Tr \big( a^\dagger a \left[ H, \rho_{\rm{s}} \right] \big)
    = g \Tr\left( a^\dagger a \left[ n_R (a+a^\dagger, \rho_{\rm{s}}) \right] \right) \\
    = g \langle n_R \left[ a^\dagger a, a+a^\dagger \right] \rangle 
    = g \langle n_R (a^\dagger-a) \rangle
\end{aligned}
\end{equation}
from the interacting term and analogously to \eqref{equ:Lindblad_dissipative_current} from the jump operator $L_\alpha=a$ (neglecting the influx due to the low bath temperature)
\begin{equation}
\begin{aligned}
    \Tr\left( a^\dagger a a\rho_{\rm{s}} a^\dagger - \frac12 a^\dagger aa^\dagger a\rho_{\rm{s}} 
    -\frac12 \rho_{\rm{s}} a^\dagger aa^\dagger a \right)\\
    = \langle a^\dagger a^\dagger a a \rangle - \frac12 \langle a^\dagger a a^\dagger a \rangle - \frac12 \langle a^\dagger a a^\dagger a \rangle
    = - \langle a^\dagger a \rangle.
 \end{aligned}
\end{equation}
Therefore, the energy current into the resonator reads
\begin{equation}
    I^E_\text{ph}
    = -i\omega_0 g \left\langle n_R (a^\dagger - a) \right\rangle
    -2\kappa\omega_0 \left\langle a^\dagger a \right\rangle,
\end{equation}
which is the expression given by \eqref{eq:energy_current_Lind1}-\eqref{eq:energy_current_Lind2} discussed in Sec.~\ref{sec:energy_current}.

In Fig.~\ref{fig:energy_curren_Lind}, we show the energy current from Lindblad calculations, comparable to the energy dissipation from the electronic system explored in Sec.~\ref{sec:Thermoelectric_current} in Fig.~\ref{fig:energy_dissipation_grid}. As expected, we find the same qualitative behavior of the energy current for both methods. However, the perturbation theory underestimates the energy current as compared to the master equation. This is consistent with the observations for the particle current made in Fig.~\ref{fig:particle_current}.

\section{Finite-bias Lindblad master equation}
\label{app:PERLind_comparision}

\begin{figure}[h]
    \centering
    \includegraphics[width=\linewidth]{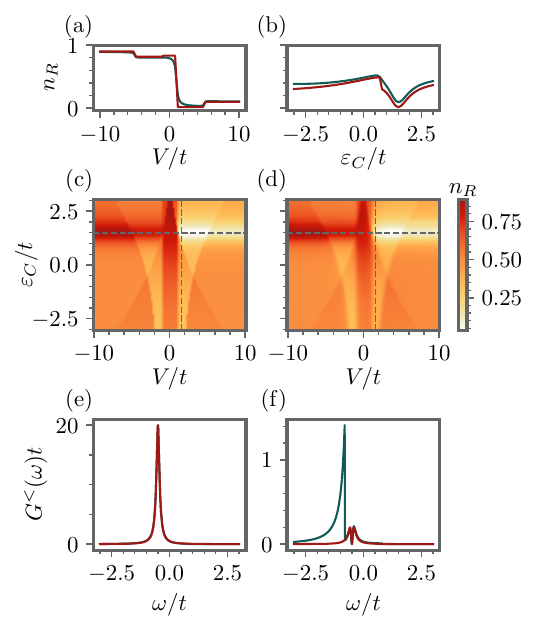}
    \caption{Occupation of the right QD calculated using PERLind for the spin-polarized TTD (panel (c) and red lines in (a), (b), (e) and (f)) and Green functions ((d) and green lines in (a), (b), (e) and (f)). Gridlines in (c)/(d) show the values of $\varepsilon_C/V$ corresponding to the cuts in (a) and (b). In (e)/(f) the LL/RR component of the lesser Green function is shown for parameters corresponding to the intersection between the gridlines in (c)/(d). Other parameters as in Table~\ref{tab:my_label}.}
    \label{fig:finite_bias_comparison}
\end{figure}

In order to assess the validity of using a finite bias voltage in the Lindblad calculation, we compare expectation values and correlation functions for the spin-polarized TTD obtained using PERLind, and Green functions respectively. This analysis is performed for $g=0$ and $U=0$, where the Green functions give the exact result. Figs.~\ref{fig:finite_bias_comparison} (a)-(d) show the expectation value of $n_R$ for the, spin-polarized TTD obtained using PERLind and the exact Green functions respectively. Although Figs.~\ref{fig:finite_bias_comparison} (c) and (d) show good agreement between the two methods, the cuts in panels (a) and (b) reveal a discrepancy, which is most prevalent close to $V=0$ and disappears for $V/2>\varepsilon_C,$ as the large-bias limit is reached. We also note that as the temperature is increased, the characteristic timescale associated with the frequency dependence of the Fermi function decreases, and good agreement between the two methods is obtained for all values of $V$. In the high-temperature limit, the validity of the Markov approximation is again ensured. 

In Fig.~\ref{fig:finite_bias_comparison} (e)-(f) an example of $G^<_{LL/RR}$ is shown for reference. The spectral function is independent of bias and is not shown. We see that the qualitative behavior of frequency-independent quantities, such as the occupation, are reasonably well reproduced by PERLind. On the other hand, frequency-dependent quantities, such as the lesser Green function, where the Markovianity of the reservoirs is important, can deviate significantly from the exact Green function results.
\FloatBarrier

\bibliographystyle{apsrev4-2}
\bibliography{main}

\end{document}